\documentclass[a4paper,12pt]{article}
\pdfoutput=1

\usepackage{amssymb,amsmath,bm}
\usepackage{a4wide}
\usepackage{color}
\usepackage{slashed}
\usepackage{cite}
\usepackage{stackengine}
\usepackage{hyperref}
\usepackage{units}
\usepackage{relsize}
\usepackage[latin1]{inputenc}
\usepackage{amsfonts}
\usepackage{lscape}
\usepackage{amsthm}
\usepackage{booktabs}
\usepackage{array}
\usepackage{rotating}
\usepackage{float}
\usepackage{multirow}
\usepackage{graphicx,rotating,amsmath,multirow,xcolor,colortbl,xcolor}
\usepackage{hyperref,pdflscape,slashed}
\usepackage{upgreek}
\usepackage{bbold}
\usepackage[normalem]{ulem}

\usepackage{tikz}
\usetikzlibrary{arrows,shapes.misc,decorations.markings,decorations.pathmorphing,positioning,intersections}
\tikzset{cross/.style={cross out, draw=black, minimum size=2*(#1-\pgflinewidth), inner sep=0pt, outer sep=0pt},
cross/.default={1.5mm}}
\tikzset{mydash/.style={dashed, dash pattern=on 4pt off 5pt}}
\tikzset{
  vertex/.style={draw,shape=circle,fill=black,minimum size=3pt,inner sep=0pt},
  cross/.style={cross out, draw=black,thick, minimum size=6pt, inner sep=0pt, outer sep=0pt},
  external/.style={inner sep=2pt},
  plabel/.style={inner sep=2pt},
  blob/.style={circle,fill=black!20,minimum size=0.7cm,draw,thick},
  whiteblob/.style={circle,fill=white,minimum size=1.0cm,draw,thick},
  effective/.style={rectangle,fill=black!20,minimum size=0.5cm,draw,thick},
  vev/.style={shape=vev,draw,inner sep=2pt,thick},
  mass/.style={shape=cross,draw,thick},
  rscalar/.style={dashed,thick},
  mfermion/.style={thick},
  scalar/.style={postaction={decorate}, decoration={markings,mark=at position .55 with {\arrow{latex}}},dashed,thick},
  ooscalar/.style={postaction={decorate}, decoration={markings,mark=at position .7 with {\arrow{latex}}},dashed,thick},
  fermion/.style={postaction={decorate}, decoration={markings,mark=at position .75 with {\arrow{latex}}},thick},
  majfermion/.style={postaction={decorate}, decoration={markings,mark=at position .7 with {\arrow{latex}}},thick},
  oofermion/.style={postaction={decorate}, decoration={markings,mark=at position .85 with {\arrow{latex}}, mark=at position .35 with {\arrowreversed{latex}}},thick},
  iifermion/.style={postaction={decorate}, decoration={markings,mark=at position .35 with {\arrowreversed{latex}}, mark=at position .85 with {\arrow{latex}}},thick},
  gaugeboson/.style={decorate, decoration={snake},thick},
  gluon/.style={decorate, decoration={coil,amplitude=4pt, segment length=5pt},thick},
  photon/.style={decorate, decoration={snake},thick},
  dashdot/.style={dash pattern=on .4pt off 3pt on 4pt off 3pt,thick}}

\definecolor{rosso}{cmyk}{0,1,1,0.4}
\definecolor{rossos}{cmyk}{0,1,1,0.55}
\definecolor{rossoc}{cmyk}{0,1,1,0.2}
\definecolor{blu}{cmyk}{1,1,0,0.3}
\definecolor{blus}{cmyk}{1,1,0,0.6}
\definecolor{bluc}{cmyk}{1,1,0,0.1}
\definecolor{verde}{cmyk}{0.92,0,0.59,0.25}
\definecolor{verdec}{cmyk}{0.92,0,0.59,0.15}
\definecolor{verdes}{cmyk}{0.92,0,0.59,0.4}
\definecolor{Gray}{gray}{0.95}

\font\tenrsfs=rsfs10 at 12pt
\font\sevenrsfs=rsfs7
\font\fiversfs=rsfs5
\newfam\rsfsfam
\textfont\rsfsfam=\tenrsfs
\scriptfont\rsfsfam=\sevenrsfs
\scriptscriptfont\rsfsfam=\fiversfs
\def\mathscr#1{{\fam\rsfsfam\relax#1}}

\usepackage[small,bf]{caption}
\hypersetup{colorlinks,bookmarksopen,bookmarksnumbered,linkcolor=blue!90!black,pdfstartview=FitH,urlcolor=black,citecolor=green!60!black}

\newcommand{\bea}{\begin{eqnarray}}
\newcommand{\eea}{\end{eqnarray}}
\newcommand{\beq}{\begin{equation}}
\newcommand{\eeq}{\end{equation}}

\numberwithin{equation}{section}

\usepackage{soul}

\begin{document}
	
{\hfill CERN-TH-2021-125}

\vspace{2cm}

\begin{center}

\boldmath

{\textbf{\LARGE Cosmological Relaxation\\ \medskip through the Dark Axion Portal}}

\unboldmath
	
\bigskip
	
\vspace{0.4 truecm}
	
{\bf Valerie Domcke},$^{a,b}$ {\bf Kai Schmitz},$^{a}$ {\bf Tevong You}$^{a,c,d}$\\[5mm]
	
{$^a$\it CERN, Theoretical Physics Department, Geneva, Switzerland}\\[2mm]
{$^b$\it EPFL, Institute of Physics, Lausanne, Switzerland}\\[2mm]
{$^c$\it DAMTP, University of Cambridge, Cambridge, UK}\\[2mm]
{$^d$\it Cavendish Laboratory, University of Cambridge, Cambridge, UK}\\[2mm]
	
\vspace{2cm}
	
{\bf Abstract }

\end{center}

\begin{quote}
The dark axion portal is a coupling of an axion-like particle to a dark photon kinetically mixed with the visible photon. We show how this portal, when applied to the relaxion, can lead to cosmological relaxation of the weak scale using dark photon production. The key backreaction mechanism involves the Schwinger effect: As long as electroweak symmetry is unbroken, Schwinger production of massless Standard Model fermions, which carry dark millicharges, suppresses the dark photon production. Once the electroweak symmetry is broken, the fermions acquire mass and the suppression is lifted. An enhanced dark photon dissipation then traps the relaxion at a naturally small weak scale. Our model thus provides a novel link between the phenomenological dark axion portal, dark photons, and the hierarchy problem of the Higgs mass. 
\end{quote}

\thispagestyle{empty}

\vfill


\newpage
{\hypersetup{linkcolor=black}\tableofcontents}


\section{Introduction}

Following the discovery of a Higgs boson~\cite{CMS:2012qbp,ATLAS:2012yve} compatible with Standard Model (SM) expectation, the lack of new physics at the TeV scale has prompted a diversification of the experimental and theoretical programme of fundamental physics. Experimentally, searches for light new physics have extended from the QCD axion~\cite{Peccei:1977hh,Weinberg:1977ma,Wilczek:1977pj,Kim:1979if,Shifman:1979if,Dine:1981rt,Zhitnitsky:1980tq}, well motivated by the strong-CP problem, to encompass a wide variety of axion-like particles~\cite{Choi:2020rgn}, dark photons~\cite{Fabbrichesi:2020wbt,Caputo:2021eaa} and dark sector fermions interacting through portal operators~\cite{Darme:2020ral}, as well as many more feebly interacting candidates~\cite{Agrawal:2021dbo}, to name just a few examples whose motivations have typically been more phenomenological. On the theoretical side, cosmological solutions involving light new physics are being explored to reconcile the hierarchy problem of the Higgs mass with an apparent separation between the weak scale and heavy new physics~\cite{Dvali:2003br,Dvali:2004tma,Graham:2015cka,Arkani-Hamed:2016rle,Arvanitaki:2016xds,Herraez:2016dxn,Geller:2018xvz,Cheung:2018xnu,Giudice:2019iwl,Kaloper:2019xfj,Dvali:2019mhn,Strumia:2020bdy,Csaki:2020zqz,Arkani-Hamed:2020yna,Giudice:2021viw, TitoDAgnolo:2021nhd}. 

In models of cosmological relaxation~\cite{Graham:2015cka,Espinosa:2015eda,Hardy:2015laa,Batell:2015fma,Marzola:2015dia,Evans:2016htp,Hook:2016mqo,You:2017kah,Evans:2017bjs,Batell:2017kho,Ferreira:2017lnd,Tangarife:2017rgl,Davidi:2017gir, Fonseca:2017crh,Son:2018avk,Fonseca:2018xzp,Davidi:2018sii,Wang:2018ddr,Gupta:2019ueh,Fonseca:2019aux,Ibe:2019udh,Kadota:2019wyz,Fonseca:2019lmc}, an axion-like particle, the so-called relaxion, scans the Higgs mass in the early universe from its naturally large value down to the observed weak-scale value, where it is trapped. For this to occur without fine tuning by hand requires the scanning to stop due to a backreaction triggered at the weak scale. In the original proposal~\cite{Graham:2015cka}, this was due to a Higgs-dependent periodic potential contribution. However, such a backreaction on the relaxion potential is problematic as the potential's dependence on the vacuum expectation value (vev) of the Higgs implies confinement of a new strong sector with electroweak charges close to the weak scale (alternatively, sterile neutrinos can be involved in the backreaction mechanism~\cite{Davidi:2018sii}). This motivated particle production as an alternative backreaction mechanism for trapping the relaxion, first proposed in Refs.~\cite{Hook:2016mqo,You:2017kah} (see also~\cite{Choi:2016kke,Tangarife:2017rgl} for other applications of particle production in relaxation). To trigger the dissipation at the weak scale, Ref.~\cite{You:2017kah} required inflation to end at that point. More minimally, Ref.~\cite{Hook:2016mqo} used the fact that electroweak gauge bosons obtain their masses from the Higgs mechanism to allow dissipation when they become sufficiently light, after scanning in the broken phase from a large Higgs vev down to a small Higgs vev. However, to avoid dissipation during scanning required a photophobic relaxion~\cite{Craig:2018kne, Fonseca:2020pjs}. Moreover, we point out here that an additional effect must be taken into account: the Schwinger production of SM fermions~\cite{Heisenberg:1935qt,Heisenberg:1936nmg,Schwinger:1951nm}. When the Schwinger effect is active, it suppresses the necessary exponential gauge boson dissipation.

In this work, we show how the dark axion portal~\cite{Kaneta:2016wvf} to a dark photon, named after the analogous axion portal coupling to SM photons, provides all the ingredients necessary for a viable mechanism of cosmological relaxation using particle production (from here on, we will use the term ``axion" to refer to the broader class of axion-like particles with the dark axion portal). Our mechanism works as follows. During scanning, the Higgs vev is stabilized at zero, so the SM fermions are massless and are efficiently produced through the Schwinger mechanism due to their millicharges under the dark $U(1)$ gauge group. This suppresses dark photon production, and the relaxion has sufficient kinetic energy to overcome its periodic potential barriers and continue slow-rolling. As the relaxion scans past the critical point, electroweak symmetry is broken by the Higgs vev and the fermions become massive. Since Schwinger suppression is lifted, the dark photon dissipation is enhanced and the relaxion loses its kinetic energy to become trapped by the periodic potential barriers. This mechanism benefits from the fact that, relative to the visible photon, the dark photon couples more strongly to the relaxion and more weakly to the SM fermions. It is the combination of these two factors that boosts dark photon production and makes the dissipation efficient enough to trap the relaxion.

The new feature that enables this relaxion mechanism is the Schwinger effect. This non-perturbative production of fermions had been investigated in the context of axion inflation in Refs.~\cite{Domcke:2018eki,Domcke:2019qmm}, and plays a role in reheating in the relaxion model of Ref.~\cite{Tangarife:2017rgl}. In our case, the Schwinger effect is an intrinsic part of the weak-scale backreaction that occurs at the critical point. Remarkably, our relaxation mechanism does not require the introduction of new physics dependent on the Higgs vacuum expectation value, since the inevitable presence of SM fermions already provides an existing SM source of backreaction. A dark axion and a kinetically mixed dark photon are all the ingredients necessary for cosmological relaxation to occur. This provides a motivation from naturalness for the dark axion portal, which had previously been introduced purely phenomenologically. 

This link between naturalness and the dark axion portal singles out a characteristic region of parameter space that that can be probed---and is highly constrained---by astrophysical and cosmological observations. Our model is therefore testable by searches for an axion interacting through the dark axion portal in the keV to MeV range. In particular, future probes of dark radiation may discover a dark relaxion portal contribution to $N_{\rm eff}$, the effective number of neutrino species in the early Universe. 

This paper is organised as follows. In the next Section we review cosmological relaxation models and set our notation. In Section~\ref{sec:darkrelaxion}, we introduce our dark relaxion model and discuss its coupling to the dark photon kinetically mixed with the visible photon. The Schwinger effect is computed in Section~\ref{sec:schwinger}, providing in particular analytical expressions for the gauge friction as a function of the fermion mass. We place phenomenological constraints on the dark relaxion in Section~\ref{sec:pheno}, before concluding in Section~\ref{sec:conclusion}. Appendix~\ref{app:thermalisation} discusses the possible thermalisation of different sectors during the relaxation process, whereas Appendix~\ref{app:thermaldarkrelics} derives the abundance of thermal relics in the late Universe. Appendix~\ref{app:inmedium} focusses on in-medium effects for the visible photon.


\section{Cosmological relaxation of the weak scale}\label{sec:review}
\subsection{Relaxation with vev-dependent periodic potential}

\begin{figure}
\centering
\includegraphics[width = 0.45 \textwidth]{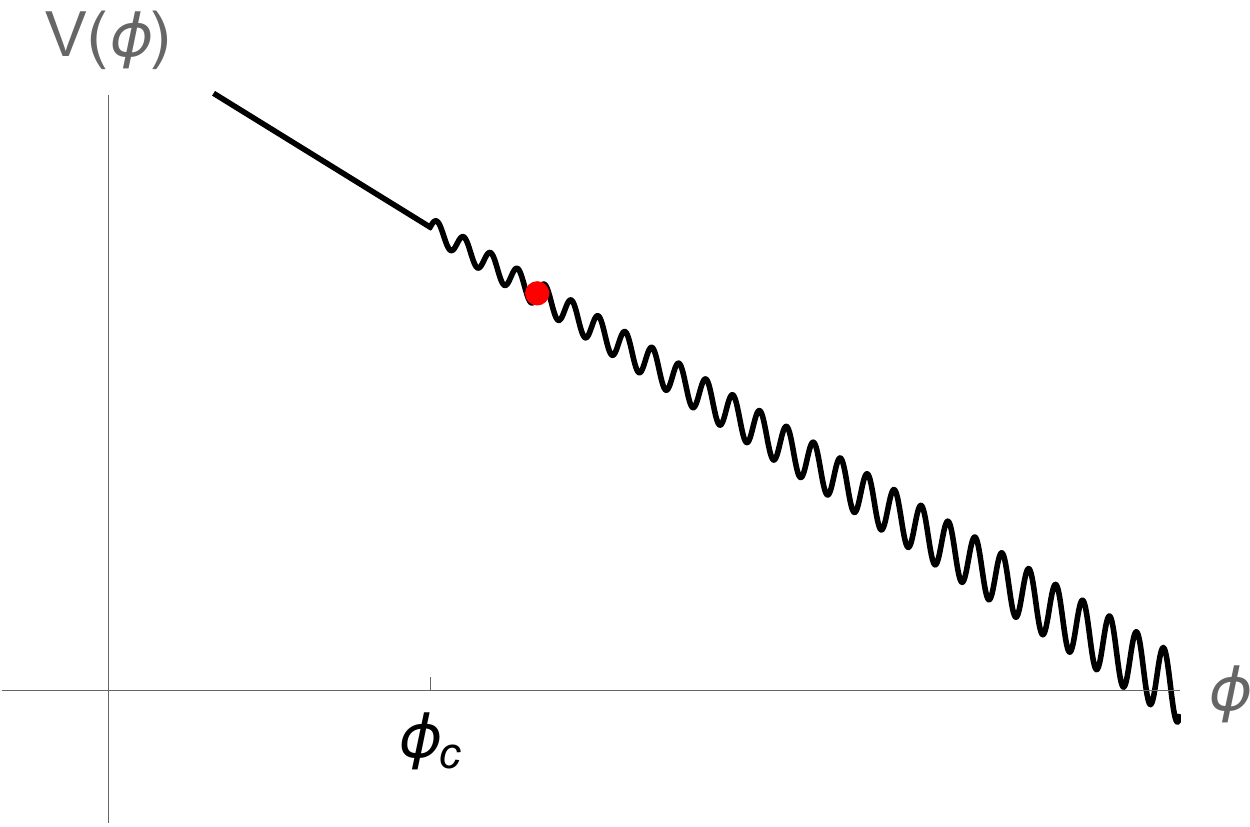} \hfill
\includegraphics[width = 0.45 \textwidth]{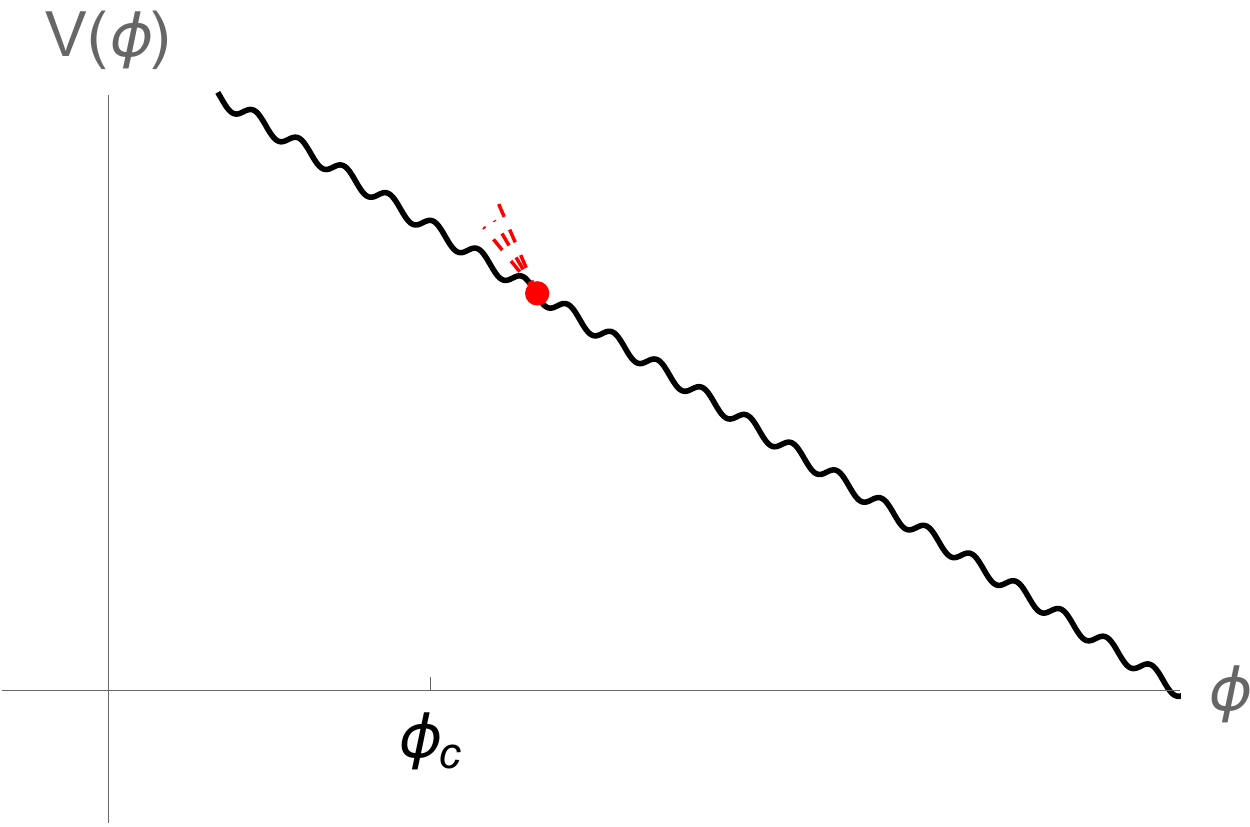}
\caption{Cartoon sketch of the relaxion potential in the model with a vev-dependent periodic potential (left) and with particle production (right). The vev-dependent periodic potential height grows linearly or quadratically with $\phi$ after passing the critical point $\phi_c = -M/\epsilon$, trapping the relaxion as shown by the red point. In the particle production mechanism, particle production is triggered after the critical point, indicated by dashed red lines, which introduces extra friction to trap the relaxion.}
\label{fig:potential}
\end{figure}

The original model of cosmological relaxation~\cite{Graham:2015cka} is described by the Lagrangian
\begin{equation}
\label{eq:GKRlagrangian}
\mathcal{L}\supset (M^2 + \epsilon M\phi)|h|^2 +\epsilon M^3 \phi + \text{...} + \Lambda_p^{4-n}v^n \cos\left(\frac{\phi}{f_p}\right) \, ,
\end{equation}
where $h$ is the Higgs field, $M$ represents the effective field theory (EFT) cut-off of the SM and $f_p$ is the relaxion decay constant entering in the periodic potential. We neglected ${O}(1)$ factors and assume masses and scales to be related by implicit $\mathcal{O}(1)$ couplings. The relaxion $\phi$ enjoys the usual discrete shift symmetry of axion-like particles that ensures the potential is periodic once generated by confinement in some strongly coupled sector at the scale $\Lambda_p$. This confinement must be proportional to the Higgs vev $v$, hence $\Lambda_p^{4-n}v^n$ where $n=1,2$ depending on the strong dynamics responsible, which can be either QCD or some new physics that also depends on $v$. In addition to the periodic potential, there are explicit shift-symmetry-breaking terms parametrised by $\epsilon$, where in Eq.~\eqref{eq:GKRlagrangian} we have kept only the leading term in a Taylor expansion of some general potential $V(\epsilon \phi)$. These can originate, for example, from clockwork-like UV completions~\cite{Choi:2015fiu,Kaplan:2015fuy,Giudice:2016yja}. From an EFT perspective, $\epsilon$ is a dimensionless parameter that can be technically naturally small since the discrete shift symmetry is restored in the limit where it is zero. This relaxion potential is illustrated in the left panel of Fig.~\ref{fig:potential}.


Through cosmological evolution, this extension of the SM Lagrangian can address the little hierarchy problem between the present-day value of the Higgs mass at the weak scale $v$ and the scale $M$ of new physics at which the Higgs mass takes its natural value, that we assume here to be at least $M \gtrsim \mathcal{O}(100)$ TeV. The $V(\epsilon \phi)$ potential generates a slope for $\phi$ along which, starting at a point $\phi \lesssim -M/\epsilon$ in the unbroken electroweak phase, it slow-rolls during inflation in the early universe. As it rolls, $\phi$ scans the effective Higgs quadratic term. We then see that despite the original Higgs mass being naturally large at the EFT cut-off, it can be made dynamically small through cosmological evolution. Crucially, there is a backreaction trapping $\phi$ when the effective Higgs quadratic term goes from positive to negative and $v \neq 0$. Since the height of the periodic potential grows with the vev $v$, it acts as a ``speed bump'' barrier preventing further evolution of $\phi$ along the slope when 
\begin{equation}
	\epsilon M^3 \simeq \frac{\Lambda_p^{4-n}v^n}{f_p} \, ,
\end{equation}
where $n=1,2$ depending on the strongly coupled sector responsible for the periodic potential. This relation naturally sets $v \ll M$.

The main disadvantage of this model is the $v$-dependence of the periodic potential term. If the periodic potential is linearly dependent on $v$ then it must either be due to QCD, which re-introduces the strong-CP problem the QCD axion was originally supposed to solve, or it requires new physics at the weak scale. If the periodic potential depends quadratically on $v$ then the sector responsible can be decoupled to higher scales, but a second scalar is then necessary to relax barriers preventing the relaxion from rolling~\cite{Espinosa:2015eda}. This leads us to consider alternative sources of backreaction. 


\subsection{Relaxation with particle production}


Particle production can occur rather generically in axion dynamics. In particular, this can happen through the axion coupling to a Chern-Pontryagin density. Since such couplings respect the classical shift symmetry of the axion, their effects can be naturally large. The relevant terms of the relaxion Lagrangian become
\begin{equation}
	\mathcal{L}\supset (M^2 + \epsilon M\phi)|h|^2 + \epsilon M^3 \phi + \text{...} + \Lambda_p^4 \cos\left(\frac{\phi}{f_p}\right) + \frac{\alpha_V}{4\pi f_V}\phi F_{\mu\nu}\tilde{F}^{\mu\nu} \, ,
	\label{eq:particleproductionlagrangian}
\end{equation}
where $\Lambda_p$ no longer depends on the Higgs vev. The field strength $F^{\mu\nu}$ depends on the particle production model. In the relaxion model of Ref.~\cite{You:2017kah}, the gauge boson was a purely dark-sector photon, while in~\cite{Hook:2016mqo} it was the photophobic combination of the electroweak gauge bosons. The coupling strength between the axion and the vector gauge bosons is parametrised by $\alpha_V/f_V$. In the next Section we will consider a kinetically mixed dark photon leading to a dark axion portal.

In models of relaxation with particle production, scanning happens with sufficient kinetic energy to overcome the periodic potential barriers, which are now independent of the Higgs vev. This potential is illustrated in the right panel of Fig.~\ref{fig:potential}. During scanning, gauge boson dissipation is sub-dominant. After passing the critical point, a backreaction enhances the dissipation and traps the relaxion once its kinetic energy is sufficiently reduced for the periodic potential to prevent further evolution. This is described by the equation of motion for the homogeneous relaxion field,
\begin{equation}
	\ddot{\phi} + 3H \dot{\phi} - \epsilon M^3 + \frac{\alpha_V \left< EB \right> }{\pi f_V} + \frac{\Lambda_p^4}{f_p} \sin\left(\frac{\phi}{f_p}\right)= 0 \, ,
\end{equation}
where $\left< EB\right>$ is the expectation value of the field strength's electric and magnetic fields produced by the $\phi$ coupling to the Chern-Pontryagin density, and the dots denote derivatives with respect to cosmic time $t$. We assume that the Hubble parameter $H$ is dominantly sourced by a separate inflation sector, providing a quasi de-Sitter background. The gauge field friction proportional to $\left< EB \right>$ must be sub-dominant during the scanning phase and enhanced after reaching the critical point. 

Neglecting the Schwinger effect, there is an exponential solution to the production of gauge boson modes~\cite{Turner:1987bw,Garretson:1992vt,Anber:2006xt}, 
\begin{equation}
	\frac{\alpha_V}{\pi f_V}\langle EB \rangle \simeq \frac{\alpha_V}{\pi f_V}\frac{I}{\xi_V^4}e^{2\pi\xi_V} H^4 \, ,
	\label{eq:EBexp}
\end{equation}
where $I \sim 10^{-4}$ and we defined
\begin{equation}
	\xi_V \equiv \frac{\alpha_V \dot{\phi}}{2\pi f_V H} \, . 
\end{equation}
For $\xi_V \gtrsim 2$, $\langle E B \rangle$ increases as the relaxion velocity $\dot \phi$ increases, justifying the notion of a friction force. This activation of gauge friction
was previously used to trigger the change in regimes at the critical point~\cite{Hook:2016mqo,You:2017kah}. However, care must be taken to include the Schwinger effect of fermions in calculating gauge boson production. Light SM fermions, in particular the electron, will suppress the exponential production of electroweak gauge bosons.

In our setup, we make use of the Schwinger suppression in reverse: during scanning, it inhibits gauge boson production; after reaching the critical point, where the quadratic term in the Higgs potential becomes negative and the fermions acquire $v$-dependent masses, the suppression is lifted. The subsequent enhancement of the gauge field production then traps the relaxion.
In the next Section we will describe our dark relaxion setup for implementing this mechanism using the dark photon.


\section{Dark relaxion portal to the dark photon}\label{sec:darkrelaxion}


The relaxion in our model is a dark axion coupling to a dark photon $X$. The dark photon can have a small mass $m_X \lesssim 10^{-15}$~eV to evade existing bounds (see Sec.~\ref{subsec:portal_constraints}), or can be treated as massless. In the first case, $m_X > 0$, we assume the dark photon mass to be generated via the St\"uckelberg mechanism, as mass generation via the Higgs mechanism might introduce another hierarchy problem in the dark sector. The dark photon naturally mixes with the visible sector photon $A$ through a small vacuum kinetic mixing parameter $\theta_D \ll 1$, 
\begin{align}
	{\cal L } \supset &  - \frac{1}{4} A'^{\mu \nu} A'_{\mu \nu} - \frac{1}{4} X'^{\mu \nu} X'_{\mu \nu} - \frac{\theta_D}{2} A'^{\mu \nu} X'_{\mu \nu}  + \frac{1}{2} m_X^2 X'_\mu X'^\mu \nonumber \\
	& \, + \frac{\alpha_D}{4 \pi f_D} \phi X'_{\mu \nu} \tilde X'^{\mu \nu}   + \sum_i   i \bar \psi_i  \gamma^\mu(\partial_\mu + i e Q_{i} A'_{\mu}) \psi_i  
	\label{eq:basis0} \, ,
\end{align}
where $\psi_i$ denote the 4-component SM fermion spinors with electromagnetic charges $Q_i$. The prime superscript on the gauge bosons in Eq.~\eqref{eq:basis0} indicates that we are in a basis with non-diagonal kinetic terms. The coupling strength between the relaxion and the dark photon is parametrised by the inverse of the relaxion decay constant, $\alpha_D/f_D$. We note that in the parameter range of interest the Hubble parameter is smaller than the QCD confinement scale, $H < \Lambda_\text{QCD} \sim 100$~MeV, and QCD confinement induces a small breaking of electroweak symmetry where the pions form the longitudinal components of the massive $Z$ and $W$ gauge bosons (see e.g.~Ref.~\cite{Contino:2010rs} for a detailed discussion). Consequently, the relevant massless SM gauge boson is the photon, even in the unbroken phase, and the relevant light degrees of freedom are the charged leptons, $i = \{e, \mu, \tau\}$. For simplicity we will nevertheless refer to the $v=0$ phase as the ``unbroken" phase (although the Higgs vev is not exactly zero) and take the broken phase to mean the usual Higgs-induced electroweak symmetry breaking.

After diagonalising to the unique basis in which both the mass matrix and the canonical kinetic terms are diagonal, we obtain up to ${\cal O}(\theta_D^2)$ corrections
\begin{align}
	{\cal L } \supset &  -\frac{1}{4} A^{\mu \nu} A_{\mu \nu} - \frac{1}{4} X^{\mu \nu} X_{\mu \nu}  + \frac{1}{2} m_X^2 X_\mu X^\mu   \nonumber \\
	& \, + \frac{\alpha_D}{4 \pi f_D} \phi X_{\mu \nu} \tilde X^{\mu \nu}   +  \sum_i  i  \bar \psi_i \gamma^\mu(\partial_\mu + i e Q_{i} A_{\mu} -  i \theta_D e Q_{i} X_\mu) \psi_i    \,.
	\label{eq:basis1} 
\end{align}
This diagonalisation is obtained by rotating only the visible photon but not the dark photon, with $A'_\mu = A_\mu - \theta_D X_\mu/\sqrt{1 - \theta_D^2}$ and $X'_\mu = X_\mu /\sqrt{1 - \theta_D^2}$. For $m_X = 0$ the two mass eigenstates become degenerate. In the absence of the axion dark photon coupling, the kinetic mixing can then be rotated away by a suitable field redefinition, rendering it unphysical~\cite{Fabbrichesi:2020wbt}. However, the coupling to the axion field results in a physical distinction of the two massless gauge fields; one couples to the axion and the other does not. In this case, the kinetic mixing cannot be simply rotated away and the physics of the massless dark photon is uniquely obtained as the $m_X \rightarrow 0$ limit. Here we consider the more general case and allow $m_X$ to vary, though our mechanism does not depend sensitively on its value so long as it is light enough to evade experimental constraints, arising in particular from the resonant conversion of cosmic microwave background (CMB) light to dark photons.

Coupling the relaxion through this dark photon portal, as opposed to the coupling to SM photons, boosts the gauge friction in a two-fold way: firstly, the SM leptons, whose masses $m$ depend on the Higgs vev $v$, only carry millicharges $\theta_D Q_i$ with respect to the dark photon gauge field, so the Schwinger production of SM fermions in a dark photon background is more suppressed, as we will see. Secondly, the dark photon axion decay constant $f_D$ is less constrained than its SM counterpart $f$ such that the relaxion--vector coupling can be stronger. 

We now consider the effect of the backreaction at the weak scale on the evolution of $\phi$ through its coupling to the dark photon. The equation of motion is
\begin{equation}
	\ddot{\phi} + 3H \dot{\phi} - \epsilon M^3 + \frac{\alpha_D \left< EB \right> }{\pi f_D} + \frac{\Lambda_p^4}{f_p} \sin\frac{\phi}{f_p}= 0 \, .
	\label{eq:relaxion_eom}
\end{equation}
where $\left< EB\right>$ is the expectation value of the dark $U(1)_D$ electric and magnetic fields. We assume the relaxion to start with some value $\phi < -M/\epsilon$ during inflation, where it scans the Higgs mass with vev $v=0$. Because the Higgs is very heavy for most of this scanning phase, the Schwinger production of Higgs quanta during relaxation is negligibly small. The SM leptons, on the other hand, are massless (since $H < \Lambda_{\rm QCD}$ we shall only consider the leptonic degrees of freedom) and so the usually explosive gauge boson production is suppressed by Schwinger production of fermions~\cite{Domcke:2018eki}. In this $v=0$ phase, we require the kinetic energy of the relaxion to be sufficiently large to overcome the potential barriers, $\dot \phi^2/2 > \Lambda_p^4$.
 
Once the relaxion scans past the critical point, $\phi > -M/\epsilon$, electroweak symmetry is broken and the Schwinger suppression is removed by the leptons acquiring mass~\cite{Domcke:2019qmm}. In the $v \neq 0 $ phase, the $\langle EB \rangle$ friction term becomes significant and allows the slow-roll kinetic energy to be dissipated efficiently. We parametrise the relative strength of dissipation with respect to the driving force of the slow-roll potential by the following ratios,
\begin{equation}
	\kappa_0 \equiv \frac{\alpha_D \left< EB\right>_0}{- \pi f_D V_{,\phi}} \ll 1 \quad , \quad 
	\kappa_v \equiv \frac{\alpha_D \left< EB\right>_v}{- \pi f_D V_{,\phi}} \gg 1 \, ,
	\label{eq:kappa0v}
\end{equation}
where the subscripts $0$ and $v$ denote the unbroken and broken phases, respectively, and to good approximation $V_{,\phi} \simeq -\epsilon M^3$. The enhancement factor of $\langle EB \rangle$ in going from the unbroken to broken phase can then be characterised by
\begin{equation}
	\kappa \equiv \frac{\langle EB \rangle_v}{\langle EB \rangle_0} = \frac{\kappa_v}{\kappa_0} \, .
	\label{eq:kappa}
\end{equation}
This must be sufficiently large to avoid fine-tuning the necessary change in dynamical regimes before and after the critical point. In the next Section we will calculate this enhancement factor $\kappa$, taking into account the Schwinger effect. 


\section{Schwinger production of fermions} 
\label{sec:schwinger}

We now turn to the effect of Schwinger production, considering the SM leptons $\psi_i$ with masses $m_i$ which carry millicharges $\theta_D Q_i$ under the dark photon gauge group. The presence of strong electric and magnetic fields induces non-perturbative particle production, as is well known from the Schwinger effect~\cite{Heisenberg:1935qt,Heisenberg:1936nmg,Schwinger:1951nm}. 

As demonstrated in Refs.~\cite{Domcke:2018eki,Domcke:2019qmm} in the context of axion inflation, this fermion production leads to an induced  current, $J_\psi^\mu = \sum_i \theta_D Q_{i} J_i^{\mu, \,\text{ind}} = \sum_i \theta_D Q_i \bar \psi_i \gamma^\mu \psi_i$,  which significantly inhibits the gauge field production. In particular, any stationary solution for the gauge field energy density, $\partial_t (E^2 + B^2) = 0$, is subject to the constraint equation
\begin{equation}
\label{eq:constraint}
E^2 + B^2 - \xi E B + \frac{e \, \theta_D E}{2 H} \sum_i  Q_{i} J^\text{ind}_i = 0 \,,
\end{equation}
with $E$ and $B$ denoting the absolute value of the dark gauge fields, $e = 0.3$ 
is the electric charge around the eV scale, $\xi$ parametrises the transfer of energy from the relaxion to the gauge fields,
\begin{equation}
	\xi \equiv \frac{\alpha_D \dot{\phi}}{2\pi f_D H} \, ,
	\label{eq:xi}
\end{equation}
and $J^\text{ind}_i$ denotes the absolute value of the  induced current parallel to the electric field,
\begin{align}
	e \, \theta_D  Q_{i} J_i^\text{ind} =  \frac{(e \, \theta_D |Q_i|)^3}{6 \, \pi^2} \frac{E B}{H} \coth\left(\frac{\pi B}{E} \right) \exp \left( - \frac{\pi m_i^2}{e \, \theta_D |Q_{i}| E} \right) \,.
	\label{eq:Jind}
\end{align}

This expression assumes  that the induced current can build up efficiently to the value given in Eq.~\eqref{eq:Jind}, which in particular requires that the diffuse fermion motion is subdominant compared to the motion induced by the acceleration in the electric field. We verify in Appendix~\ref{app:thermalisation} that this is indeed the case as long as $\theta_D \gtrsim 10^{-6}$. This expression moreover neglects the dissipation into SM photons, sourced by the moving charges forming the induced current. The formation of a large-scale non-thermal photon field configuration is impeded by the Schwinger effect, which limits the gauge field growth in the visible sector more severely than in the dark sector. We thus expect the resulting energy in the visible photon field to be at most comparable with its dark counterpart, at most leading to an ${\cal O}(1)$  correction of the dark photon abundance.  As we argue in App.~\ref{app:thermalisation}, we expect this visible photon field to retain its non-thermal distribution. 
We neglect in-medium effects for the SM photon due to the presence of charged fermions here for simplicity, which could potentially have a significant impact but are beyond the scope of the present discussion. 
More definite statements would require solving the highly non-linear coupled relaxion-dark photon-fermion-SM photon system, which is a challenging problem beyond the scope of the current paper.  

Due to the dependence of the induced current~\eqref{eq:Jind} on the gauge fields, the equation of motion for the gauge fields, taking into account the backreaction through the induced current, is non-linear and can no longer be decomposed into independent Fourier modes. Instead, we can exploit the observation that for $J_\psi \neq 0$, Eq.~\eqref{eq:constraint} forms a closed contour in the $E$\,--\,$B$ plane. This allows, for any given constant value of $\xi \gtrsim 2$, to determine an absolute upper bound for the value of $EB$ entering the gauge friction term in Eq.~\eqref{eq:relaxion_eom} (see Refs.~\cite{Domcke:2018eki,Domcke:2019qmm} for more details). 

\begin{figure}
	\centering
	\includegraphics[width = 0.48 \textwidth]{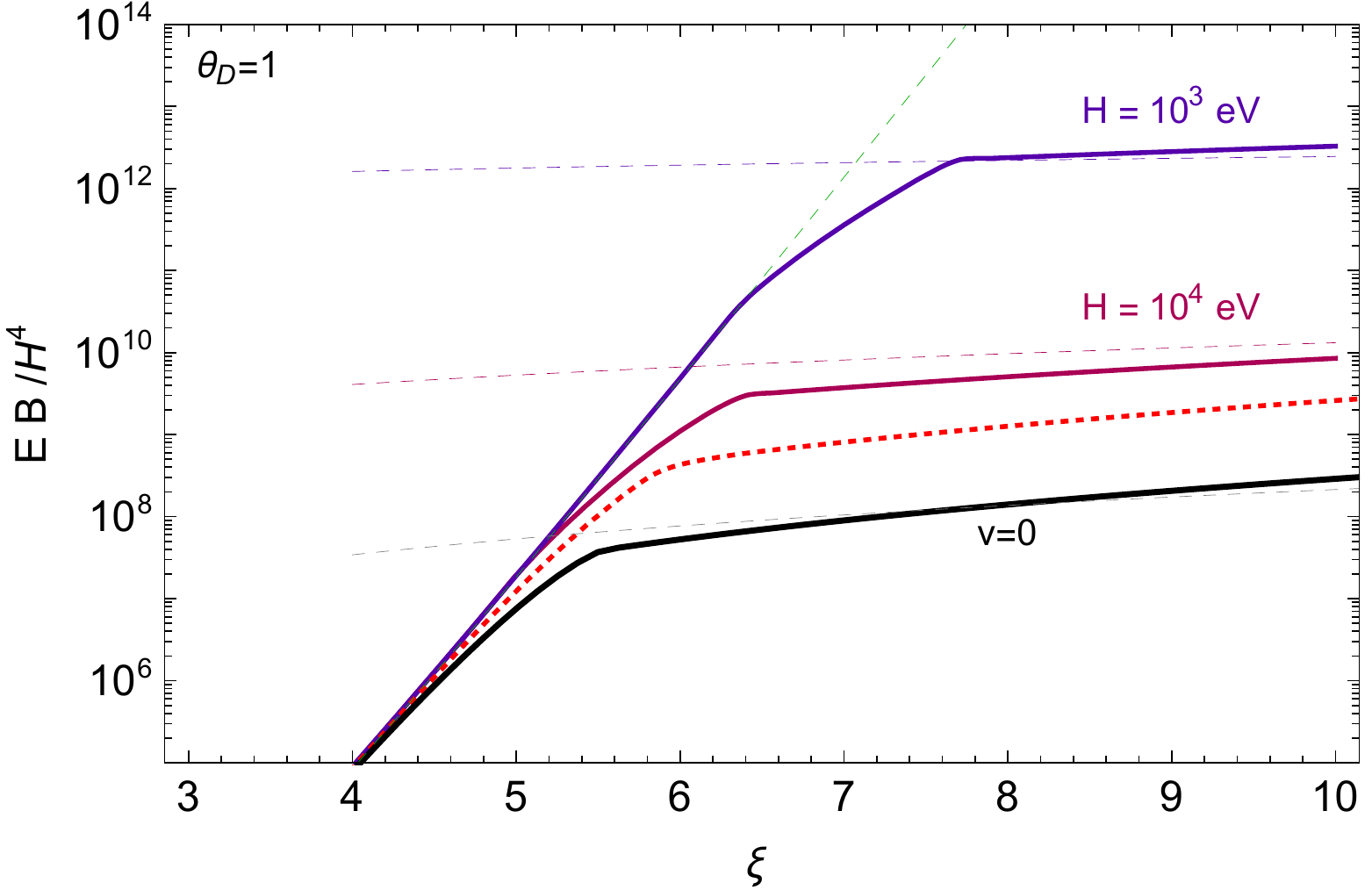} \hfill
	\includegraphics[width = 0.48 \textwidth]{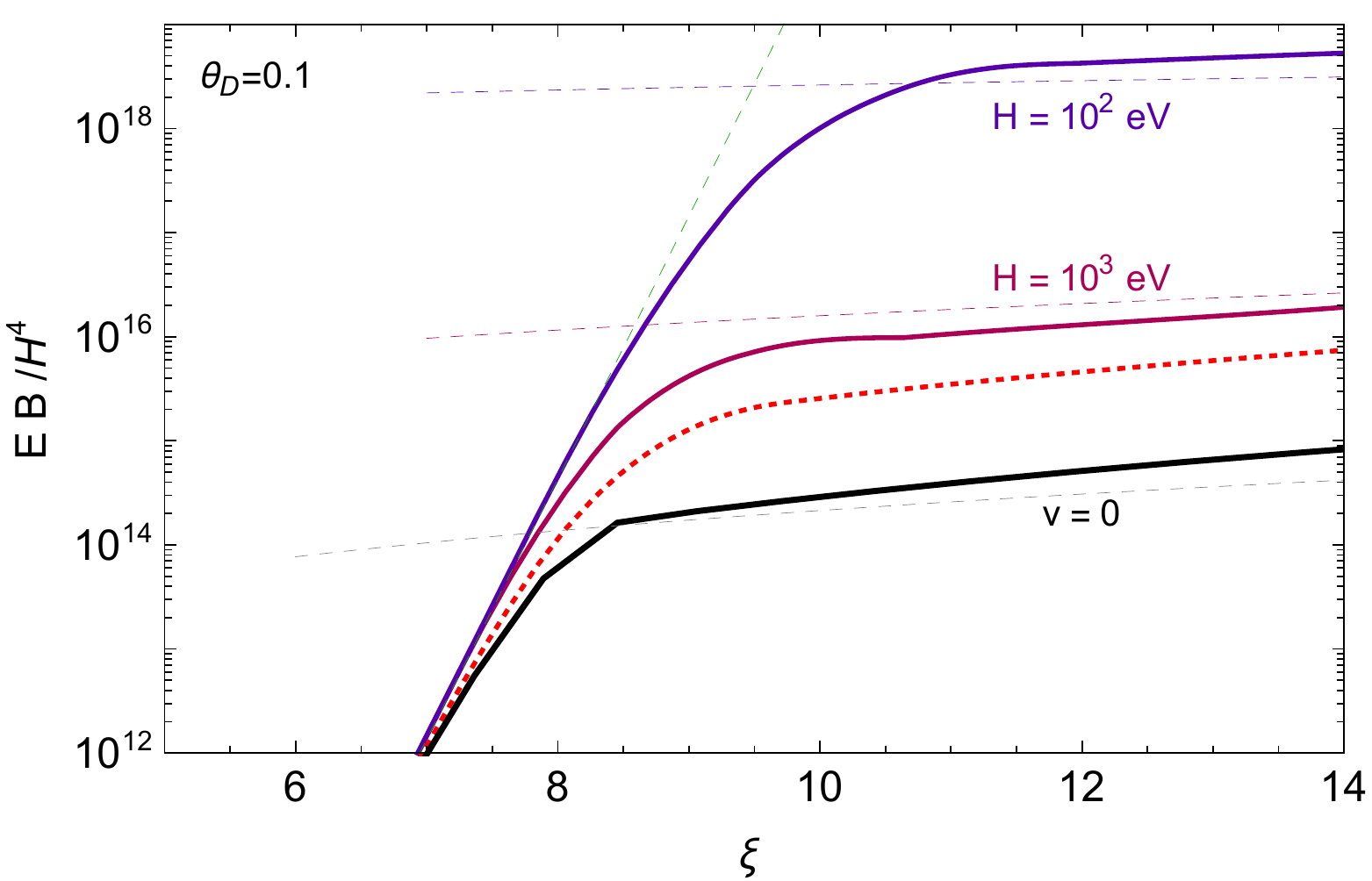}
	\caption{Upper bound on gauge field production including backreaction from SM leptons with unit charge (left) and millicharge $\theta_D = 0.1$ (right). The black contour corresponds to a vanishing Higgs vev, including the backreaction of all three massless SM charged leptons on the gauge field production. The coloured contours correspond to a Higgs vev of 246 GeV for different values of the Hubble parameter $H$, with the dotted red line indicating the limit $H \rightarrow \infty$. Here we include only the electron as the lightest charged SM particle. The dashed lines show our analytical approximations, given by Eq.~\eqref{eq:EBexp} (green), \eqref{eq:EBm0} (grey) and \eqref{eq:EB} (purple) with $\beta = 1$ and $\tilde \beta = 3$.}
	\label{fig:EB}
\end{figure}

This is depicted in Fig.~\ref{fig:EB} which shows the upper bound on $\left< EB \right>/H^4$ as a function of $\xi$ for a vanishing Higgs vev $v=0$ (black solid line) and finite Higgs vev $v=246$ GeV, with different values of $H = 10^4$~eV and $10^3$~eV (red and purple solid lines), and $\theta_D=1$ and 0.1 in the left and right panels, respectively. We see the increase in $\langle EB \rangle$ as we go from a vanishing Higgs vev (black) to a finite Higgs vev (purple), in particular for large values of $m_e/H$, which is at the core of our mechanism. Decreasing the mixing angle $\theta_D$ (right panel) increases the overall amount of dark photon production, which is crucial to generate a significant amount of friction once the Higgs acquires a vev. 

The backreaction of the gauge fields on the relaxion equation of motion is negligible when $v = 0$, even for relatively large values of $\langle EB\rangle /H^4$, since the driving force in the relaxion equation of motion $\propto V_{,\phi}$ dominates over the gauge friction term. When $v \neq 0$ backreaction effects are important; this is precisely what causes the trapping of the relaxion. The detailed dynamics of this process are expected to be complicated, since around the critical point the $\xi$-parameter drops abruptly. The gauge friction will follow with a short delay (see Ref.~\cite{Domcke:2020zez} for a related analysis in the context of axion inflation). However, for our relaxion mechanism we only require the kinetic energy to be dissipated sufficiently to be trapped by the periodic potential, so it will not be necessary to track the dynamics in detail (see {\it e.g.}\ Refs.~\cite{Choi:2016kke,Banerjee:2021oeu} for related studies albeit without the inclusion of the Schwinger effect). In the remainder of this section we will derive analytic expressions for the parametric dependence of the gauge friction in the unbroken and broken phases as defined in Eq.~\eqref{eq:kappa0v}, respectively, and then discuss the implications for our relaxion model.


\subsection{Unbroken phase}

As long as the Higgs vev is stabilised at zero, all SM leptons are massless and the induced current is given by Eq.~\eqref{eq:Jind} with $m_i = 0$ and the subscript $i$ running over the charged SM leptons $e$, $\mu$ and $\tau$. Inserting this into Eq.~\eqref{eq:constraint}, we obtain an upper bound for the dark gauge field production in the regime where the fermion production is efficient,
\begin{equation} 
	\frac{\langle E B \rangle}{H^4} = \beta \, \frac{\xi^2}{c^2}  \quad \text{for } m = 0  \quad \text{with } 
	c =  \frac{e^3 \theta_D^3}{4 \pi^2} = 7 \cdot 10^{-4} \, \theta_D^3\,,
	\label{eq:EBm0}
\end{equation}
with $\beta = 1$ indicating the upper bound for $E = B$, $\beta = {\cal O}(1)$ for $E \neq B$ and $\beta \lesssim {\cal O}(1)$ if these upper bounds are not saturated. The dashed gray line in Fig.~\ref{fig:EB} shows a comparison of this expression with the numerical results. With this, we can determine the necessary condition to ensure that the relaxion velocity is sufficient to overcome the potential barriers,
\begin{equation}
\label{eq:rolling_cond}
\frac{1}{2} \dot \phi^2 > \Lambda_p^4 \,.
\end{equation}

The relaxion velocity is obtained by algebraically solving the slow roll equation,
\begin{equation}
 3 H \dot \phi - \epsilon M^3 + \frac{\alpha_D}{\pi f_D} \langle E B \rangle \simeq 0 \,,
 \label{eq:sr_simple}
\end{equation}
with the gauge field configuration~\eqref{eq:EBm0} depending on the relaxion velocity through $\xi$ (defined in Eq.~\eqref{eq:xi}). This gives 
\begin{align}
 \dot \phi_0 \simeq \frac{6 c^2 \pi^3 f_D^3}{\beta \alpha_D^3 H} \left( \sqrt{1 + \frac{\beta \alpha_D^3 \epsilon M^3 }{9 c^2 \pi^3 f_D^3}} - 1 \right)\,,
 \label{eq:dphi_unbroken}
\end{align}
with the subscript 0 indicating the unbroken phase.

We also require $\xi \gtrsim \mathcal{O}(10)$, depending on $\theta_D$, to ensure that we are in a regime where the fermion backreaction on the gauge field production is relevant, see Fig.~\ref{fig:EB}. With
\begin{equation}
	\xi \simeq 20 \left(\frac{5\cdot 10^{-9}\text{ GeV}}{H}\right)^2\left(\frac{10^4\text{ GeV}}{f_D/\alpha_D}\right)\left(\frac{M}{10^5\text{ GeV}}\right)^3\bigg(\frac{\epsilon}{10^{-25}}\bigg) \,,
\end{equation}
this can be taken as a lower bound on the slope $\left|V_{,\phi}\right| \simeq \epsilon M^3$. For smaller values of $\xi$, the gauge field production grows exponentially with $\xi$ irrespective of the fermion masses.

As an aside, we note that the suppression of the gauge boson friction due to the presence of light fermions presents a previously unaccounted for obstacle to the mechanism proposed in~\cite{Hook:2016mqo}. A key ingredient for relaxation during inflation in~\cite{Hook:2016mqo} is the large gauge friction which stops the motion of the relaxion field once the vev of the Higgs field becomes sufficiently small. However, including Schwinger suppression\footnote{The gauge boson produced in this scenario is the photophobic linear combination of Abelian gauge boson of the SM, hence the Schwinger suppression will be dominated by neutrinos which are much lighter than the Hubble scale. This is why we use $\kappa_0$ instead of $\kappa_v$ in the following estimate.},
 $ \kappa_0 \simeq 100 \; \xi^2 H^4 /(M^2 f_V \epsilon)$,
which for the benchmark value presented in~\cite{Hook:2016mqo} for the case of relaxation during inflation evaluates to $\kappa_0 \simeq 10^{-28}/\xi^2$, in contrast to the efficient stopping condition $\kappa_0 \gg 1$. Here we have set $\theta_D = 1$ since in \cite{Hook:2016mqo} the gauge boson is the photophobic linear combination of the SM Abelian gauge bosons.

This estimate relies on an approximately constant value of $\xi$ which is certainly not the case around the critical point. Let us thus take a closer look at the relevant time scales. The tachyonic instability of a single mode proceeds with a typical time scale $\Delta t_\text{pp} \sim H M_P f_V^3/\dot \phi^3 $, which is the estimate used in Ref.~\cite{Hook:2016mqo}. However, a significant change of the gauge field background $\langle E B \rangle$ (for massless Abelian gauge fields without fermion production)  requires a longer timescale $\Delta t_{\langle E B \rangle} \sim \ln(\xi^2/2)/ H $.\footnote{
For slowly varying $\xi$, a mode of physical wave number $k$ obtains a maximally negative effective square mass at $k/H = \xi$. On the other hand, the integrand of $\langle E B \rangle$ assumes its maximal contribution at $k/H = 2/\xi$. Correspondingly, $\langle E B \rangle$ is dominated by modes whose maximal growth period originates from about $\ln(\xi^2/2)$ e-folds earlier. See~\cite{Domcke:2020zez} (and in particular Eq.~(17) therein) for a detailed discussion and numerical simulations in the absence of fermions.
} The inverse of the Schwinger production rate yields $\Delta t_\text{schw} \sim H^{-1}/\xi^2$ and is much shorter (note that the benchmark value presented in~\cite{Hook:2016mqo} corresponds to a value of $\xi \sim 10^{10}$ before reaching the critical point, much larger than the values discussed in this paper), $\Delta t_\text{schw} \ll \Delta t_{\langle E B \rangle}$. This suggests that Schwinger production instantaneously reacts to the increasing gauge field background and hence the large values of the gauge friction necessary to stop the relaxion in~\cite{Hook:2016mqo} are efficiently prevented by Schwinger suppression. A more detailed study of the parameter space of this setup including numerical simulations of the dynamics around the critical point are required for a final verdict on this point.


\subsection{Broken phase}

Once the Higgs obtains a vacuum expectation value, all fermions obtain masses proportional to the Higgs vev $v$, implying an exponential suppression of the induced current~\eqref{eq:Jind}. The dominant remaining contribution now comes from the electrons, the lightest SM particles charged under $U(1)_\text{em}$. The resulting upper bound on the dark gauge fields becomes
\begin{align}
	\frac{\langle EB \rangle}{H^4} = \tilde \beta  \, \frac{d^2}{w^2} \quad \text{with} \quad w \simeq W_0 \left(  \frac{\tilde c \, d}{\xi} \right) \,, \quad \tilde c = \frac{ e^3 \theta_D^3}{12 \pi^2} = 2 \cdot 10^{-4} \;\theta_D^3 \,, \quad d = \frac{\pi m_e^2}{ e \, \theta_D H^2} \,,
	\label{eq:EB}
\end{align}
where the product logarithm $W_0(z)$ denotes the solution of $z = W_0 \exp(W_0)$. As above, $\tilde \beta = 1$ indicates the upper bound on the helicity obtained for the special case $E = B$, with $\tilde \beta \lesssim 10$ otherwise. This is illustrated in Fig.~\ref{fig:EB}. We note that since it is by no means guaranteed that the actual value of the helicity generated by the rolling relaxion field saturates this upper bound, we may also consider values of  $EB = \tilde \beta \, d^2/w^2$ with $\tilde \beta < 1$.

This increase in $\langle E B \rangle$ implies an increase in the gauge friction (see Eq.~\eqref{eq:sr_simple}) and thus a rapid drop in the relaxion velocity. The details of this transition can be rather complicated, and a detailed numerical investigation of this problem is beyond the scope of this paper. Instead, we impose two simple and conservative requirements for rapidly stopping the relaxion in the unbroken phase. 

Firstly, we require the gauge friction immediately after the phase transition to dominate over the driving force of the slow roll potential,
\begin{equation}
\label{eq:exclPh2}
\kappa_v = \frac{\alpha_D}{\pi f_D} \frac{\tilde{\beta}\,d^2/w^2}{\epsilon\,M^3}\, H^4 > 1\,,
\end{equation}
where $\langle E B \rangle$ in Eq.~\eqref{eq:EB} is evaluated using the relaxion velocity at the phase transition, $\dot \phi = \dot \phi_0$. Inserting this into Eq.~\eqref{eq:sr_simple} yields 
\begin{equation}
\label{eq:phiphi0}
 \dot \phi \simeq \dot{\phi}_0\left(1 - \kappa_v\right) \,,\quad \dot{\phi}_0 \simeq \frac{\epsilon M^3}{3 H} \,,
 \end{equation}
again indicating that $\kappa_v > 1$ implies completely stopping the relaxion due to this gauge friction. Strictly speaking, this is not a necessary condition since even a small velocity decrease could be sufficient to ensure that the relaxion is trapped by the barriers of the periodic potential. This, however, requires a fine-tuning of the energy scales, which we want to avoid. Note also that in reality the gauge friction decreases as soon as the relaxion velocity decreases, leading to a more complicated dynamics of the coupled system. 

Secondly, we require the would-be equilibrium value of the relaxion kinetic energy that would be reached in the absence of the periodic potential barriers to be smaller than the energy of the potential barriers,
\begin{equation}
\label{eq:trappingcond}
\frac{1}{2} \dot \phi_v^2 < \Lambda_p^4 \,,
\end{equation}
where $\dot \phi_v$ is determined by algebraically solving Eq.~\eqref{eq:sr_simple}, but now with the gauge field configuration given in Eq.~\eqref{eq:EB}, which in turn depends on $\phi_v$.
There are two distinct regimes, parametrised by the argument of the product logarithm, $z \equiv {\tilde c}\,d/\xi$. 

When $z \gg 1$, the electron mass efficiently suppresses fermion production, and we can approximate $W_0(z) \simeq \ln(z)$. Therefore, with 
\begin{equation}
	 z = \frac{e^2 \theta_D^2}{12\pi\xi}\left(\frac{m_e}{H}\right)^2 \simeq 60 \left(\frac{10}{\xi}\right) \left(\frac{\theta_D}{10^{-3}}\right)^2\left(\frac{\text{ eV}}{H}\right)^2 \gg 1 \,,
\end{equation}
we find
\begin{align}
\label{eq:dphi_broken}
\dot \phi_v \simeq \frac{\epsilon M^3}{3 H} - \frac{\tilde \beta \alpha_D d^2 H^4}{3 \pi \ln^2(z) f_D H} < \frac{\epsilon M^3}{3 H} \,.
\end{align}
Note that this is an implicit equation for $\dot \phi_v$, since on the right-hand side, $z$ is a function of $\dot \phi_v$. However, in this regime, the right-hand side depends only logarithmically on the relaxion velocity, such that the change in velocity occurring when transitioning into the broken phase does not have large impact.  In the following, we will therefore simply use the slow-roll velocity in the unbroken phase, $\dot{\phi}_0 \simeq \epsilon M^3/\left(3H\right)$, in order to evaluate the logarithmic factor in $\left<EB\right>$ right after the onset of the broken phase. In this approximation, the two velocities in Eqs.~\eqref{eq:phiphi0} and \eqref{eq:dphi_broken} are identical.

In this regime, switching on the Higgs vev enhances the gauge friction, as parametrised by the ratio $\kappa = \kappa_v/\kappa_0$, by a factor
\begin{align}
	\left. \kappa \right|_{z \gg 1} \simeq 3 \cdot 10^6 \, \bigg( \frac{\tilde \beta \ln(z)}{\beta \xi^4} \bigg) \left( \frac{\text{eV}}{H} \right)^4 \left( \frac{\theta_D}{10^{-3}} \right)^4 \,.
\end{align}
This large enhancement of the gauge field friction enables a natural trapping of the Higgs vev close to the SM value. 

On the other hand, for $z \ll 1$, the electric fields are so large that the suppression induced by the electron mass becomes insignificant. In this case,  $W_0(z) \simeq z$, and we recover the expression~\eqref{eq:dphi_unbroken} for the relaxion velocity with $\beta \mapsto \tilde \beta$ and $c \mapsto \tilde c$. In this case, the enhancement factor of the gauge friction in going from unbroken to broken electroweak symmetry tends to a constant ratio, 
\begin{equation}
	\left. \kappa \right|_{z \ll 1} \simeq \frac{c^2}{\tilde c^2} \frac{\dot \phi_0^2}{\dot \phi^2} \simeq  9 \, ,
\end{equation}
In this regime, the change induced in the relaxion velocity when crossing into the broken phase is not very pronounced, justifying neglecting it when estimating the gauge friction in the broken phase. We see that when $z \ll 1$ the proposed mechanism can only be implemented at the cost of some tuning. Note, moreover, that as $z \rightarrow 0$, the muon and tau masses eventually also become negligible, implying $\kappa \rightarrow 1$. We will therefore place a limit on the parameter space requiring $z > 1$ to avoid this fine-tuned regime.

The condition $z > 1$  sets a lower bound on $\theta_D^2 m_e/H$, and hence for fixed values of $H$ and $\theta_D$ determines a lower bound for the Higgs vev. As we will see in the next section, combining all the contraints on this mechanism limits the parameter space to a rather small window around $z = \mathcal{O}(1)$. In this sense, the selection criteria for the Higgs vev can be phrased as
\begin{align}
 z \sim 1 \quad \Rightarrow \quad v \sim (100~\text{GeV}) \left( \frac{\xi}{10} \right)^{1/2}  \left( \frac{10^{-3}}{\theta_D} \right)  \left( \frac{H}{\text{eV}} \right)\,.
\end{align}

\section{Phenomenology of the dark relaxion portal}
\label{sec:pheno}

In this Section we consider constraints on the dark relaxion. We first list the bounds from the relaxion mechanism itself, before showing how the remaining parameter space for our model is constrained by the phenomenology of the dark axion portal~\cite{Kaneta:2016wvf,Choi:2018dqr,deNiverville:2018hrc,Choi:2018mvk,Choi:2019jwx,Arias:2020tzl,Deniverville:2020rbv,Co:2021rhi,Ge:2021cjz,Hook:2021ous}. Relaxion phenomenology has been studied more generally in Refs.~\cite{Choi:2016luu,Flacke:2016szy,Fonseca:2018kqf,Banerjee:2018xmn,Fonseca:2019lmc,Banerjee:2020kww,Fonseca:2020pjs,Barducci:2020axp,Banerjee:2021oeu, Balkin:2021wea}.

\subsection{Constraints from the dark relaxion mechanism}

Our free parameters are the Hubble scale $H$, the EFT cut-off $M$, the shift-symmetry breaking parameter $\epsilon$, the dark sector's periodic potential $\Lambda_p$ and its associated decay constant $f_p$, the dark axion's decay constant $f_D$ parametrising its coupling to the dark photon, and the dark photon's mixing angle $\theta_D$. The coupling $\alpha_D$ is degenerate with $f_D$, so we absorb this factor into the definition of $f_D$ (or, equivalently, we set $\alpha_D=1$). We set $\beta=\tilde{\beta}=1$, which are typically values of $\mathcal{O}(1)$ or less if the $EB$ bound is not saturated, as discussed in the previous Section. 

We require that viable points in our parameter space satisfy the following criteria: 
\begin{itemize}
	\item The relaxion's energy density is sub-dominant to the inflaton's: $H > M^2/M_p$. 
	\item Classical rolling beats Hubble quantum fluctuations: $\epsilon M^3 > H^3$.
	\item The kinetic energy in the unbroken phase must be sufficient to overcome the periodic potential barriers: $\epsilon M^3 > H \Lambda_p^2$. 
	\item The periodic potential forms local minima: $\epsilon M^3 < \Lambda_p^4 / f_p$.
	\item The distance between minima changes the weak scale by less than the observed value: $2\pi f_p < v^2/\epsilon M$.
	\item Inefficient dissipation in unbroken phase: $\dot{\phi}_0^2 > \Lambda_p^4$ with $\dot{\phi}_0$ given by Eq.~\eqref{eq:dphi_unbroken}.
	\item Efficient dissipation in going from unbroken to broken phase: $\xi > 10$, where $\xi$ is determined by the slow-roll velocity in the unbroken phase, and $\dot{\phi}_v^2 < \Lambda_p^4$ with $\dot{\phi}_v$ given by Eq.~\eqref{eq:dphi_broken}. To avoid fine-tuning we also require $\kappa_v > 1$.  
	\item Require a large change in the gauge friction between the unbroken and broken phase to avoid fine-tuning: $z = \frac{e^2 \theta_D^2}{12\pi\xi}\left(\frac{m_e}{H}\right)^2 > 1$, with $\dot{\phi}$ in $\xi$ conservatively taken to be the slow-roll velocity when entering the broken phase. 
	\item Avoid relaxion fragmentation, using the criteria given by Eq.~4.42 of Ref.~\cite{Fonseca:2019lmc}.
\end{itemize}
A benchmark set of parameter values that satisfies all these constraints (as well as the phenomenological constraints listed in the next subsection) is given in Table~\ref{tab:benchmark}. The relaxion mass for this benchmark point is given by 
\begin{equation}
    m_\phi = \frac{\Lambda_p^2}{f_p} = 9 \text{ keV} \left(\frac{\Lambda_p}{3\cdot 10^{-2}\text{ GeV}}\right)^2 \left(\frac{10^2\text{ GeV}}{f_p}\right) \, .
\end{equation}
The dark sector's decay constants and confinement scales are relatively low, indicating that there must be relatively light new physics in the dark sector. However, their only interaction with the visible sector is through the dark axion portal. In particular, the degrees of freedom responsible for confinement and generating the periodic potential need not communicate with the visible sector at all.  

\begin{table}[t]
\begin{center}
\begin{tabular}{|c|c|c|c|c|c|c|}
    \hline
    $H$ & $M$ & $\epsilon$ & $\Lambda_p$ & $f_p$ & $f_D/\alpha_D$ & $\theta_D$   \\
    \hline
    $5\cdot 10^{-9}$ GeV & $10^5$ GeV & $10^{-25}$ & $3\cdot 10^{-2}$ GeV & $10^2$ GeV  & $10^4$ GeV & $10^{-3}$ \\
    \hline
\end{tabular}
\end{center}
\caption{Benchmark point satisfying all constraints on our relaxion model.}
\label{tab:benchmark}
\end{table}

We note that for the large values of $\xi$ required in this mechanism, the axion-assisted Schwinger effect~\cite{Domcke:2021fee} could be important. This would exponentially enhance the gauge boson production in the broken phase for
\begin{align}
	\alpha_D^2 \dot \phi^2/f_D^2 \gtrsim \frac{\pi m_e^2 p_T^2}{e \theta_D |Q| E} \,.
\end{align}
However, for momenta $p^2 \simeq e \theta_D E$ this would require $\xi \gtrsim (m_e/H)^2$, which is well outside the parameter regime discussed here.

We also have an upper bound on the number of e-foldings required to scan an $\mathcal{O}(M/\epsilon)$ field range to avoid entering the eternal inflation regime (see e.g.~\cite{Rudelius:2019cfh}),
\begin{equation}
    N_e = \left(\frac{H}{\epsilon M}\right)^2 < 8\pi^2\left(\frac{M_p}{H}\right)^2 \, .
\end{equation}
This can be written as a constraint on the Hubble parameter, 
\begin{equation}
    H < 10^{-1} \,\text{GeV} \bigg(\frac{\epsilon}{10^{-25}}\bigg)^{1/2} \left( \frac{M}{10^5 \text{ GeV}} \right)^{1/2} \, ,
\end{equation}
which is always satisfied in our parameter space. 

\begin{figure}[t]
	\centering
	\includegraphics[width = 0.45 \textwidth]{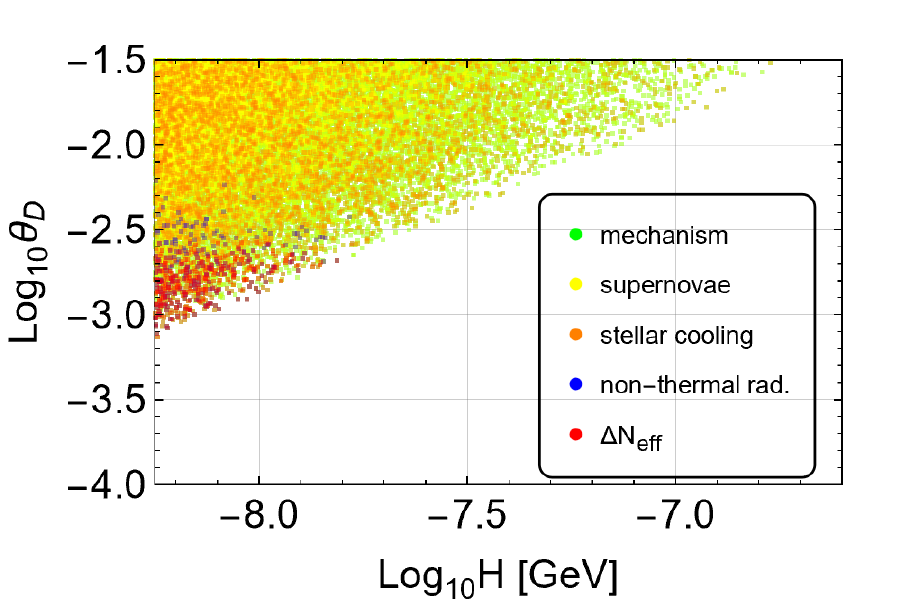}
	\includegraphics[width = 0.45 \textwidth]{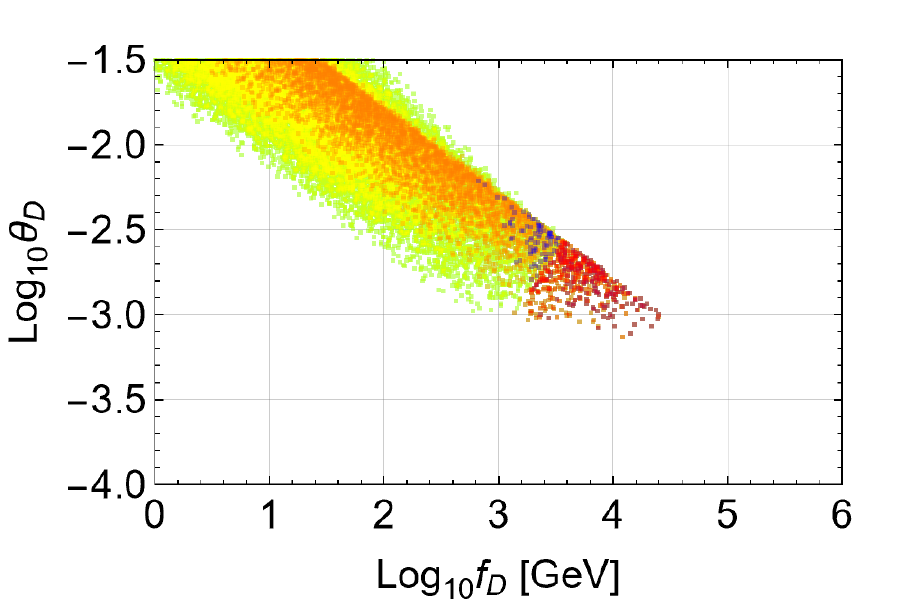} \\
	\includegraphics[width = 0.45 \textwidth]{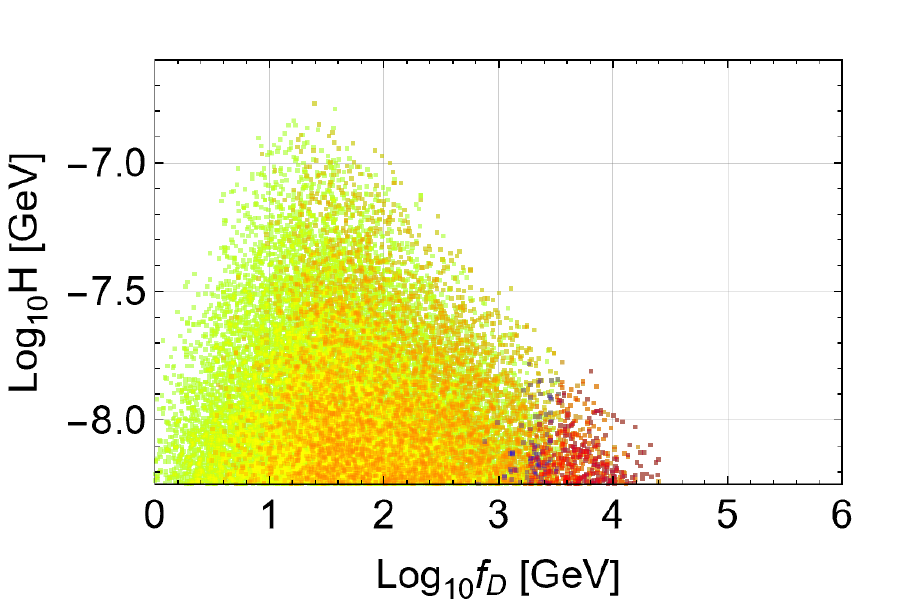}
	\includegraphics[width = 0.45 \textwidth]{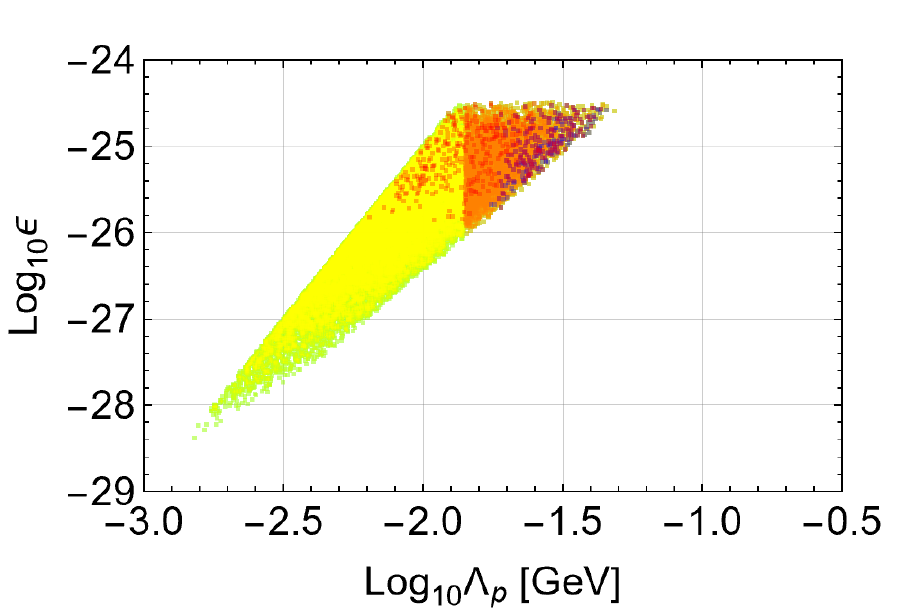}
	\caption{Parameter space scan of points that satisfy the criteria for a successful dark relaxion mechanism (green) and that additionally pass one phenomenological constraint from either supernovae (yellow), stellar cooling (orange), non-thermal dark radiation production from post-reheating rolling (blue), or dark radiation contribution to $\Delta N_\text{eff}$ from thermal scattering (red).}
	\label{fig:parameterspace1}
\end{figure}

To better understand how these constraints shape the parameter space, we scanned over the parameters $H, \epsilon, \Lambda_p, f_D$, and $\theta_D$, while fixing $M = 10^5$ GeV and $f_p = 10^2$ GeV. The result is plotted in Fig.~\ref{fig:parameterspace1}. The green dots are points that pass all the criteria above after randomly sampling 10 million points with flat priors on a log scale over the range displayed on the axes of the projection plots. Note that there is a hard lower cut-off on the Hubble scale $H \gtrsim 5\cdot 10^{-9}$ GeV from the sub-dominant energy density criteria. 

The top left plot shows that as $\theta_D$ becomes larger, we can access larger values of the Hubble parameter. This reflects the dependence of $\kappa = \left<EB\right>_v/\left<EB\right>_0$ on the ratio $\left(\theta_D/H\right)^4$ in Eq.~\eqref{eq:kappa}, which follows from the fact that $\kappa \propto c^2 d^2$, where $c\propto \theta_D^3$ and $d\propto 1/(\theta_D\,H^2)$. More precisely, larger $\theta_D$ implies larger millicharges and hence a larger induced current in the unbroken phase, see Eq.~\eqref{eq:Jind}, which in turn enhances the Schwinger suppression of $\left<EB\right>_0$. This results in an increase of $\kappa$ that can be undone by a larger value of $H$, which reduces the exponential suppression of the induced current in the broken phase, see Fig.~\ref{fig:EB}, and hence equally leads to a stronger Schwinger suppression of $\left<EB\right>_v$, thus compensating for the change in $\left<EB\right>_0$. In the top right figure, we see a strong anti-correlation between $\theta_D$ and $f_D$. This reflects the fact that for smaller values of the mixing angle the fermions have smaller millicharges, there is less Schwinger suppression, and so the coupling to dark photons inversely proportional to $f_D$ can be weaker for the same effect. In the bottom right, the correlation between $\epsilon$ and $\Lambda_p$ is due to the size of the periodic potential barriers having to be correspondingly larger for a steeper slope. Finally, in the bottom left, we see again that weakening the dissipation through a larger $f_D$ requires reducing the Hubble scale in order to also reduce the Schwinger effect in the broken phase.   

\subsection{Constraints from the dark relaxion portal \label{subsec:portal_constraints}}

\subsubsection*{Dark photon} 

Our model can be probed by searches for a kinetically mixed dark photon. However, for the purpose of the dark relaxion mechanism, the mass of the dark photons can be as light as necessary, even massless, to evade all constraints. Bounds on the dark photon mixing parameter were recently collected in Refs.~\cite{Caputo:2021eaa, Fabbrichesi:2020wbt}. We assume $m_X \lesssim 10^{-15}$~eV, which is lighter than the plasma mass of the photon after recombination. This forbids a resonant conversion between photons and dark photons, evading the CMB distortion bounds~\cite{Caputo:2020bdy}. 
Heavier masses are still possible so long as the mixing parameter can be small enough to satisfy observational constraints; the only strict upper bound is that they remain sufficiently light relative to the Hubble parameter for the dark photons to be efficiently produced.

\subsubsection*{Stellar cooling}

Stronger constraints arise from the relaxion coupling to the kinetically mixed dark photon. Various possibilities for such dark axion models have been considered in the literature, making different assumptions on the couplings $g_{\phi XX}$, $g_{\phi X\gamma}$, $g_{\phi\gamma\gamma}$ in the Lagrangian terms
\begin{equation}
	\mathcal{L} \supset g_{\phi XX}\phi X_{\mu\nu}\tilde{X}^{\mu\nu} + g_{\phi X\gamma}\phi X_{\mu\nu}\tilde{A}^{\mu\nu} + g_{\phi \gamma\gamma}\phi A_{\mu\nu}\tilde{A}^{\mu\nu} \, .
\end{equation}
Even if $g_{\phi X\gamma}$ and $g_{\phi\gamma\gamma}$ are set to zero, nevertheless there still exist bounds from {\it e.g.}\ stellar cooling due to the mixing-induced interaction between SM fermions and the relaxion. This is easily seen by going to an ``interaction" basis in which $g_{\phi X\gamma}$ and $g_{\phi\gamma\gamma}$ no longer vanish. 

Consider our starting point, Eq.~\eqref{eq:basis0}, reproduced here for convenience:
\begin{align}
	{\cal L } \supset &  - \frac{1}{4} A'^{\mu \nu} A'_{\mu \nu} - \frac{1}{4} X'^{\mu \nu} X'_{\mu \nu} - \frac{\theta_D}{2} A'^{\mu \nu} X'_{\mu \nu}  + \frac{1}{2} m_X^2 X'_\mu X'^\mu \nonumber \\
	& \, + \frac{\alpha_D}{4 \pi f_D} \phi X'_{\mu \nu} \tilde X'^{\mu \nu}   + \sum_i   i \bar \psi_i  \gamma^\mu(\partial_\mu + i e Q_{i} A'_{\mu}) \psi_i   \, .
\end{align}
If $\phi$ is an axion originating from some global $U(1)_A$ breaking and the dark photon belongs to some dark $U(1)_X$ gauge symmetry, then this EFT Lagrangian can be generated in the UV by two sets of fermions: one set charged under $U(1)_A$ and $U(1)_X$ but singlets under the SM gauge groups, so they can be responsible solely for the anomalous coupling of $\phi$ to dark photons; while another set of fermions are singlets under $U(1)_A$ but charged under $U(1)_{\rm em}$ and $U(1)_X$ so they induce kinetic mixing without contributing to a mixed axion-photon-dark photon coupling. In this case we have $g_{\phi XX} = \alpha_D/(4\pi f_D)$ and $g_{\phi X\gamma}, g_{\phi\gamma\gamma} \ll g_{\phi XX}$ suppressed by higher-order loop corrections.\footnote{The axion--vector couplings are protected by the underlying continuous shift symmetry in the massless axion limit. However, for a massive axion, they can still receive corrections proportional to the mass squared, since the mass is the parameter that introduces an extra source of explicit shift symmetry breaking.}

In Eq.~\eqref{eq:basis1}, we rotated to the mass eigenstate basis with canonical kinetic terms, which introduced a coupling of the SM fermions $\psi_i$ to the dark photon. This is a suitable basis for calculating the Schwinger effect. However, in this basis the SM matter interacts with the combination $A_\mu + \theta_D X_\mu$ and it is this combination that is effectively the visible photon relevant for stellar cooling constraints. To make this explicit, we can choose to rotate instead to an interaction basis which only removes the mixed kinetic term without modifying the interaction of the SM fermion coupling to the photon. Up to $\mathcal{O}(\theta_D^2)$, this is
\begin{align}
	{\cal L } \supset &  - \frac{1}{4} A'^{\mu \nu} A'_{\mu \nu} - \frac{1}{4} X''^{\mu \nu} X''_{\mu \nu} + \frac{1}{2} m_X^2 \left( X''_\mu X''^\mu + 2\theta_D X''_\mu A'^\mu \right) \nonumber \\
	& \, + \frac{\alpha_D}{4 \pi f_D} \phi X''_{\mu \nu} \tilde X''^{\mu \nu} + \frac{\alpha_D \theta_D}{2 \pi f_D} \phi {A^\prime}_{\mu \nu} \tilde X''^{\mu \nu}   + \sum_i   i \bar \psi_i  \gamma^\mu(\partial_\mu + i e Q_{i} A'_{\mu}) \psi_i   \, .
	\label{eq:interactionbasis}
\end{align}


\begin{figure}
\centering
\begin{tikzpicture}[node distance=1cm]
\coordinate[label=left:$\gamma$] (v1);
\coordinate[vertex, right = of v1] (v2);
\coordinate[below right = of v2, label=right:$X$] (v3);
\coordinate[above right = of v2, label=right:$\phi$] (v4);
\draw[gaugeboson]  (v1) -- (v2);
\draw[rscalar] (v4) -- (v2);
\draw[gaugeboson] (v3) -- (v2);
\end{tikzpicture}
    
\caption{Feynman diagram of the process relevant for stellar cooling constraints on the dark relaxion, where a plasmon $\gamma$ decays to the dark relaxion $\phi$ and a dark photon $X$.}
\label{fig:stellarcooling}
\end{figure}
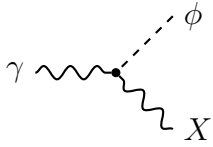


Bounds from stellar cooling can therefore be placed on our dark relaxion despite initially appearing to couple only to dark photons. In the interaction basis, we explicitly have a coupling to a photon and dark photon. Ref.~\cite{Arias:2020tzl} places the bound $g_{\phi X\gamma} = \alpha_D \theta_D/(2\pi f_D) \lesssim 3\cdot 10^{-9} \text{ GeV}^{-1}$ from horizontal-branch stars with the process shown by the Feynman diagram in Fig.~\ref{fig:stellarcooling}, which bounds the axion decay constant $f_D$ from below,
\begin{equation}
	f_D \gtrsim \frac{\alpha_D \theta_D}{6\pi}\, 10^{9} \text{ GeV} \, .
	\label{eq:fdstellarcooling}
\end{equation}
This constraint can be evaded if the relaxion mass is heavier than the plasma mass of the photon, which in practice requires $m_\phi \gtrsim 2$ keV, kinematically forbidding the decay into a relaxion final state. White dwarfs require instead $m_\phi \gtrsim 20$ keV~\cite{Hook:2021ous}, though we omit this constraint as there have been hints of an excess that may relax the bound~\cite{Giannotti:2015kwo,Giannotti:2017hny} (see, however, also Ref.~\cite{Dessert:2021bkv}). Slightly stronger limits can in principle be obtained from red giants~\cite{Kalashev:2018bra, Capozzi:2020cbu} which we do not include here as this has not yet been cast in terms of constraints on the dark axion portal. For heavier masses, bounds from supernova constraints on the dark axion coupling $g_{\phi X\gamma}$ have recently been placed in Ref.~\cite{Hook:2021ous}. 


\subsubsection*{Thermal dark relics}


Once the dark relaxion mechanism described above has ensured that the Higgs is settled to its vacuum expectation value at $\langle h \rangle \sim 246$~GeV, the standard cosmological evolution proceeds. The remaining period of vacuum energy domination must be chosen so that the cosmological perturbations in the CMB can be reproduced. This inflationary epoch strongly dilutes any relic dark sector or SM particles which are produced during the relaxation phase. At the end of inflation, the Universe can reheat up to 
\begin{equation}
	T_\text{rh} \lesssim \left( \frac{90}{\pi^2 g_*} \right)^{1/4} \sqrt{H M_P} \simeq 2.6 \cdot 10^4 \text{ GeV} \left( \frac{H}{\text{eV}} \right)^{1/2}\,,
\end{equation}
with $g_* = 106.75$ denoting the SM degrees of freedom at high temperature. A further upper bound on the reheating temperature arises from the requirement that the global symmetry featuring the relaxion as an angular degree of freedom should not be restored,
\begin{equation}
 T_{\text{BBN}} < T_\text{rh} < f_p \,,
 \label{eq:PQbound}
\end{equation}
otherwise the stabilisation mechanism for the vacuum expectation value of the Higgs vev would be lost.
In the following we will assume that the dark sector particles, i.e.\ the relaxion and the dark photon, are not directly coupled to the inflaton sector and hence are not directly produced in the reheating process.
We will further assume that any other dark sector particles, in particular the ones responsible for the barriers in the relaxion potential, decay into the relaxion and dark photon before the dark and visible sectors decouple. The relevant production mechanisms for the relaxion and dark photon are then 
through their couplings to the SM plasma as well as non-thermally through the relaxion misalignment mechanism.

From the discussion above we note that in the preferred parameter space, the dark photon is much lighter than the relaxion,
$m_X < 10^{-15}~\text{eV} \ll m_\phi$. Any relaxion abundance will thus rapidly decay into dark photons with a decay rate\footnote{ 
For non-perturbative cosmological axion decay, see \cite{Alonso-Alvarez:2019ssa}.
}
\begin{equation}
	\Gamma_{\phi \rightarrow XX} = \frac{\alpha_D^2}{64 \pi^3 f_D^2}\, m_\phi^3 \simeq (130~\text{s})^{-1} \left(\frac{10^4~\text{GeV}}{f_D/\alpha_D}\right)^2 \bigg(\frac{m_\phi}{10~\text{keV}}\bigg)^3 \,.
\end{equation}
The dark photon remains stable and ultra-relativistic throughout the evolution of the Universe, and thus contributes to dark radiation.

For $T > m_\phi, m_e$, the dominant production of dark radiation occurs through electron positron annihilation, $e^+ e^- \rightarrow A' \rightarrow X \phi$, see Ref.~\cite{Hook:2021ous} for a related discussion.\footnote{
	Any thermal dark relic production channel must necessarily contain not only the dark photon but also the relaxion, since in the absence of the relaxion the visible sector and the dark sector decouple in the interaction basis~\eqref{eq:basis0}, up to dark photon / visible photon oscillations mediated by the kinetic mixing (see e.g.~\cite{Jaeckel:2008fi}). The latter are subdominant here since we take the dark photon to be much lighter than the plasma mass of the ordinary photon at all times.
} 
If this 
interaction is efficient at high temperatures, it leads to a thermal equilibrium between the dark sector and the SM sector. Ensuring that the total dark radiation is subdominant compared to SM radiation, as imposed by the $\Delta N_\text{eff}$ constraints from big-bang nucleosynthesis (BBN) and CMB decoupling, requires this interaction to decouple before the QCD phase transition. The subsequent annihilation of the coloured degrees of freedom then reheats the SM thermal bath sufficiently to dominate over the dark radiation contribution. This requires (see App.~\ref{app:thermaldarkrelics} for details)
\begin{align}
\label{eq:Neff}
T_\text{dec,d} \simeq 10^3 \text{ MeV } g_*^{1/2}(T) \left( \frac{10^{-3}}{\theta_D} \right)^2 \left( \frac{f_D/\alpha_D}{10^4 \text{ GeV}} \right)^2    \gtrsim 100~\text{MeV}\,,
\end{align}
which, combined with Eq.~\eqref{eq:PQbound}, implies $f_p > T_\text{rh} > 100$~MeV. 
%

\subsubsection*{Non-thermal dark relics} 

A further non-thermal production channel stems from the motion of the relaxion field after reheating. If the reheating temperature is larger than the height of the potential barriers, $T_\text{rh} > \Lambda_p$, then these barriers vanish allowing the relaxion to roll down the potential $V(\phi) = - \epsilon M^3 \phi$, thereby non-perturbatively producing dark photons. As long as $T > T_\text{dec} > T_\text{BBN}$, such that the dark sector is still in equilibrium with the visible sector, this leads to an unobservable overall heating of the thermal bath. But the contribution from $T_\text{dec} > T > \Lambda_p$ will contribute to $\Delta N_\text{eff}$, leading to a further constraint on the model parameters.

Let us first conservatively assume that the entire potential energy released during this period is released into relativistic dark degrees of freedom contributing to $\Delta N_\text{eff}$. For simplicity, we will assume that that the parameter $\xi$ encoding the relaxion velocity approaches a constant value $\xi_c$ once the gauge friction becomes efficient~\cite{Choi:2016kke}.\footnote{
The coupled system of axion, gauge fields and fermions has to our knowledge never been simulated in radiation domination, we hence base this assumption on simulations performed of the simpler case of only axions and gauge fields~\cite{Choi:2016kke}. This is clearly a point that deserves further investigation.
}
With this, we can estimate the total amount of potential energy released after decoupling as
\begin{equation}
\Delta V = \int dV \left(\frac{a(t)}{a(t_{\Lambda_p})}\right)^4 = \int_{t_\text{dec}}^{t_{\Lambda_p}} \frac{dV}{d\phi} \frac{d\phi}{dt} \, dt \left( \frac{t}{t_{\Lambda_p}} \right)^2 = \frac{\pi}{2} \epsilon M^3 \frac{f_D}{\alpha_D}  \left( 1 - \frac{\Lambda_P}{T_\text{dec}} \right) \,.
\end{equation}
Satisfying the $\Delta N_\text{eff}$ bound~\cite{Pisanti:2020efz,Yeh:2020mgl} implies
\begin{equation}
 \Delta V \lesssim \Delta N_\text{eff} T^4 \lesssim 0.5 \, \Lambda_p^4 \,,
\end{equation}
which imposes an upper bound on the slope of the linear potential, 
\begin{align}
	\epsilon  < \xi_c^{-1} \, 10^{-31} \, \left( \frac{\Lambda_p}{\text{MeV}} \right)^4 \left( \frac{10^5 \, \text{GeV}}{M} \right)^3 \left( \frac{10^4 \, \text{GeV}}{f_D/\alpha_D} \right)\,.
	\label{eq:postreheatingrolling2}
\end{align}
for $T_\text{dec} \gg \Lambda_p$.\footnote{
During the process, the relaxion is displaced by $\Delta \phi \simeq 2 \pi \xi_c f_D/\alpha_D$. Note that this displacement does not interfere with the relaxation of the Higgs vev, as $\Delta \phi < v^2/(\epsilon M)$ in the parameter range of interest.
}
This simple constraint may, however, be 
overestimating the dark photon production. 
Firstly, the assumption of a constant $\xi_c$ may be incorrect, and the gauge friction may stop the rolling of the relaxion even before the barriers reappear. Secondly, for $T_\text{dec} \sim \Lambda_p$, the contribution to dark radiation is significantly suppressed. 
We thus consider the constraint~\eqref{eq:postreheatingrolling2} to be less robust than the other constraints presented in this work, though we include it to illustrate the potential size of the effect based on our rough estimates.

Once the temperature drops to $T \sim \Lambda_p$, the potential barriers re-appear. Before settling into its local minimum, the relaxion may oscillate producing relaxion (and consequently dark photon) relics via the misalignment mechanism. 
The total energy density contained in the relaxion oscillations,
\begin{align}
	\rho_\phi^\text{osc} \lesssim  m_\phi^2 f_p^2
	\simeq \Lambda_p^4\,, 
\end{align}
can be comparable to the energy of the SM thermal bath for a generic large misalignment angle. For $T_\text{dec} \lesssim \Lambda_p$, this additional entropy is absorbed by the SM thermal bath. For  $T_\text{dec} > \Lambda_p$ we require an accidentally small misalignment angle to avoid an overproduction of dark radiation.

\subsubsection*{Summary of phenomenological constraints}

In Fig.~\ref{fig:parameterspace1}, we plot the main constraints discussed above on the parameter space: 
\begin{itemize}
	\item The relaxion must be either sufficiently heavy, $m_{\phi} \simeq \Lambda_p^2/f_p > 2 \text{ keV}$, or weakly coupled, see the constraint Eq.~\eqref{eq:fdstellarcooling}, in order to evade bounds from stellar cooling in horizontal-branch stars. 
	\item To satisfy the supernova constraints, the dark relaxion's decay constant $f_D/\alpha_D$, as a function of $m_\phi$, must lie outside the region excluded by Ref.~\cite{Hook:2021ous}.
    \item The global symmetry in the dark sector from which the relaxion originates must not be restored in the thermal plasma after inflation, which sets a bound on $f_p$ from Eq.~\eqref{eq:PQbound}. 
	\item The thermal production of dark radiation must not be over-abundant, $\Delta N_{\text{eff}} < 0.5$, see Eq.~\eqref{eq:Neff}.  
	\item Negligible non-thermal dark radiation production from post-reheating rolling, see Eq.~\eqref{eq:postreheatingrolling2}.
\end{itemize}
The scan is performed as described previously, by randomly sampling 10 million points with log flat priors over the ranges shown. Points in Fig.~\ref{fig:parameterspace1} that satisfy the supernova constraints are shown in yellow. This follows the outline of the green parameter space points allowed by the relaxion mechanism since there is a degeneracy between $\theta_D$ and $f_D$ in the overall coupling that is relevant for the supernova bound. This has the effect of simply reshuffling the allowed parameter space points with only values at the boundaries breaking this degeneracy, hence the density of points thinning out at the edges. The stellar cooling constraints in orange are more severe. In particular, there is a lower bound on $\Lambda_p$ that comes from requiring a minimum relaxion mass to forbid plasmon decay. Finally, the strongest constraints come from non-thermal radiation production in post-reheating rolling (blue) and thermal radiation contribution to $\Delta N_{\text{eff}}$ (red). The points that satisfy all constraints correspond to the regions where red and blue points overlap. The lower bound on $\epsilon$ from thermal radiation is due to a tension between efficient gauge boson dissipation and efficient thermal scattering contributing to $\Delta N_{\text{eff}}$: $\epsilon$ parametrises the steepness of the potential slope, so reducing the slow-roll velocity by making the slope flatter would reduce dark photon dissipation unless the coupling is also increased by lowering $f_D$, which maintains $\xi \propto \epsilon/f_D > 10$. However, lowering $f_D$ leads to a stronger coupling which increases the thermal scattering abundance.   

\section{Conclusions}
\label{sec:conclusion}

Naturalness, based on the sound reasoning of effective field theory, has long been a guiding principle in the search for new physics. For example, the strong-CP problem motivated the QCD axion solution. Here, we proposed a solution to the Higgs mass hierarchy problem that motivates a dark axion portal to the dark photon. A central role is played by the Schwinger effect which leads to non-perturbative production of SM fermions in a dark gauge field background. Remarkably, our economical model does not require a new physics source of weak-scale backreaction since the SM fermion masses already provide the necessary trigger at the critical point. 

We investigated the parameter space of our mechanism and find that current astrophysical and cosmological constraints bound the parameter space from all directions while leaving a viable region unexplored. Cosmological probes of $\Delta N_{\text{eff}}$ place the strongest bounds from dark relic production affecting big-bang nucleosynthesis and the cosmic microwave background. It is encouraging that a corner of parameter space remains non-trivially compatible with all other constraints, such as stellar cooling or the non-thermal relic abundance from post-reheating rolling. Further study will be needed to place these constraints on firmer footing. 
In particular, 
a detailed numerical study of the fermion backreaction, including also the dissipation into SM photons and the interactions of the SM photons with the fermion plasma, is required in order to eliminate uncertainties in the dark photon spectrum. Such in-medium effects, in particular, may lead to important modifications of the dynamics discussed here. Moreover, it would be interesting to numerically solve the relevant Boltzmann equations to obtain a more accurate $\Delta N_{\text{eff}}$ constraint. This will be particularly important in the event of a future detection of dark radiation.    

\paragraph{Acknowledgments.} We thank Simon Knapen for insightful discussions on the dark photon constraints as well as Anson Hook and Kyohei Mukaida for valuable comments on the manuscript. We also thank Nayara Fonseca, Enrico Morgante, Geraldine Servant, and an anonymous referee for suggestions leading to an improved v2. T.\,Y.\ is supported by a Branco Weiss Society in Science Fellowship and partially by the UK STFC via the grant ST/P000681/1. This project has received funding from the European Union's Horizon 2020 Research and Innovation Programme under grant agreement number 796961, "AxiBAU" (K.\,S.)

\appendix
\section{Thermalisation}
\label{app:thermalisation}

Once the SM fermions are produced from the motion of the relaxion, they can interact through SM gauge interactions which are not suppressed by the small mixing angle $\theta_D$, which can lead to a thermalisation of the fermion sector. In this appendix, we discuss the conditions under which this thermalisation occurs and the consequences for the induced current and the dark gauge boson configuration. See also Refs.~\cite{Ferreira:2017lnd,Domcke:2018eki,Domcke:2019qmm} for a discussion within the SM, i.e.\ $\theta_D = 1$. 
 
\paragraph{Induced current.} In a background electric field pointing in $z$-direction, the induced current is given by the number or particles passing through the area $dA$ (orthogonal to $\hat e_z$) in a time $dt$, 
\begin{align}
	J_z = \frac{\# \text{ particles}}{dA \, dt} = n_\psi\,\frac{dz}{dt} = n_\psi \, v_z \,.
\end{align}
For relativistic particles (as will be the case in the entire parameter space of interest), the component of the velocity in $z$-direction $v_z$ can be expressed by the ratio of momentum parallel ($p_z$) and transverse ($p_\perp$) to the $z$-axis,
\begin{align}
	p_z \gg p_\perp   \quad &\Rightarrow \quad v_z \simeq 1 \quad \quad \quad \Rightarrow J_z = n_\psi \,, \\
	p_z \ll p_\perp    \quad &\Rightarrow \quad v_z \simeq p_z/p_\perp \quad \Rightarrow J_z = \frac{p_z}{p_\perp}\, n_\psi \,.
\end{align}
In other words, if the random motion of particles due to diffusion processes dominates over the acceleration induced by the electric field, the induced current is (not surprisingly) significantly suppressed. Note that Eq.~\eqref{eq:Jind} given in the main text relies on $p_z \gg p_\perp$.

We estimate $p_z$ as the momentum acquired from accelerating in the background field, $p_z \simeq e \theta_D Q E \tau$, where $\tau$ denotes the average scattering time-scale of the fermion (see below). The transverse momentum at production can be estimated as $p_\perp \sim (e \theta_D Q E)^{1/2}$~\cite{Domcke:2019qmm}; however, once the fermions thermalise (in their rest frame) a more accurate estimate is obtained from the temperature of this thermal bath, $p_\perp \sim T_\psi$, see below.

\paragraph{Fermion temperature.} Assuming the fermions thermalise, we can estimate the temperature of the resulting thermal bath as 
\begin{align}
	T_\psi^4 \simeq \frac{15}{\pi^2}\,\bar n_\psi\, \bar \omega \,,
	\label{eq:Tpsi}
\end{align}
with $\bar n_\psi$ and $\bar \omega$ denoting typical values for the fermion number density and energy~\cite{Domcke:2019qmm}, taking for simplicity unit charge $Q=1$ for all species,
\begin{align}
	\bar n_\psi = \dot{\bar n}_\psi H^{-1} = \theta_D^2 \frac{e^2}{4 \pi^2} \frac{E B}{H} \exp\left(-\frac{2 \pi B}{E}\right) \exp\left(- \frac{\pi m^2}{e \theta_D E}\right) \,, \qquad \bar \omega = (e \theta_D E)^{1/2} \,.
	\label{eq:average_values}
\end{align}
Here we have assumed that fermion production is efficient, $m^2 \lesssim e \theta_D E$, and that Landau levels with with non-vanishing transverse momentum (so-called higher Landau levels) contribute significantly to the total fermion energy. Moreover we take $B \sim E$.
If the fermions do not thermalise, the temperature $T_\psi$ nevertheless provides a useful reference value for the 'would-be' temperature of the fermion distribution.

A conservative estimate to check the thermalisation of the fermion sector is to require that the thermal scattering rate, $\Gamma_\text{th} \simeq \alpha^2 T_\psi$ be smaller than the Hubble expansion rate $H$ with $\alpha = e^2/(4 \pi)$. In this case, even an initial thermal distribution of fermions would quickly drop out of thermal equilibrium. Inserting the expression~\eqref{eq:Tpsi}, this leads to an upper bound on gauge field strength above which thermalisation is possible according to this criterion.

\paragraph{Scattering time scale.} If the fermions are not yet in thermal equilibrium, their typical scattering rate can be estimated as
\begin{align}
	\Gamma_\text{sc} = \bar n_\psi \sigma_\text{sc} \,, \qquad \sigma_{sc} = \frac{4 \pi \alpha^2}{3 \bar \omega^2}\,.
	\label{eq:Gamma_sc}
\end{align}
For $\Gamma_{sc} < H$ the fermions do not thermalise.
For $\Gamma_{sc} \gg H$, the fermion population will quickly thermalise, in which case the scattering timescale is given by $\Gamma_\text{th}^{-1}$.

\paragraph{Thermalisation within the dark gauge boson sector.} Even if the SM thermalises through the exchange of SM gauge bosons, the dark gauge bosons can still feature a non-thermal distribution. Sustaining a thermal equilibrium between fermions, dark photons and visible photons requires efficient Compton scattering involving a dark and a visible photon. The corresponding thermal scattering rate is
\begin{align}
	\Gamma_\text{th}^X \simeq  \theta_D^2 \alpha^2 T_X \,,
	\label{eq:GammaX}
\end{align}
with $T_X$ denoting the would-be temperature of the dark gauge bosons,
\begin{align}
	T_X^4 = \frac{30}{\pi^2 g_X} (E^2 + B^2) \,,
	\label{eq:TX}
\end{align}
where $g_X = 2$. As long as $\Gamma_\text{th}^X < H$, the dark gauge bosons do not thermalise.

\paragraph{Separation of scales.} The induced current in Eq.~\eqref{eq:basis1} provides a source term for a non-thermal SM photon field configuration with a characteristic wavelength set by the characteristic scale of the dark gauge field configuration, $\lambda \sim (\xi H)^{-1}$. In the basis of Eq.~\eqref{eq:basis0} the same non-thermal photon field configuration is sourced via kinetic mixing from the dark gauge fields. As long as these SM gauge fields are to good approximation homogeneous on the length scales relevant for fermion production and propagation, i.e.\ as long as there is a clear separation of scales between the macroscopic classical gauge field configuration and the microphysics associated with the fermions, the dynamics of the gauge fields is well described by a coarse-grained effective theory where the microscopic effects are captured by the induced current, 
which limits the magnitude of gauge fields, analogous to the discussion in Sec.~\ref{sec:schwinger}.
Note that this does not alter the mixing angle $\theta_D$ between the dark and the visible gauge sector. Our discussion neglects any other possible effects of fermion plasma on the long-wavelength SM photons.

These conclusions however no longer hold if efficient scattering processes shift the photon distribution function to shorter wavelengths. Below, we will present a conservative estimate for this in both the unbroken and broken phase, based on the $\theta_D = 1$ limit of Eq.~\eqref{eq:GammaX}. We note however that this is the scattering rate for a thermal photon population, whereas in the our case the large correlation length of the photon distribution implies that the momentum exchanged in a Compton scattering process $\sqrt{t}$ is small compared to total center-of-mass energy $\sqrt{s}$. From the differential Compton scattering cross section, we thus expect an additional suppression factor $t/s$ for the scattering rate of the non-thermal photon configuration. In conclusion, we will see that we can safely take photon scattering involving the low-momentum photons of the background gauge field configuration to be inefficient in the entire relevant parameter space.

\subsection{Unbroken phase}
\label{app:thermalisation-unbroken}

\paragraph{Scattering rates.} As long as the induced current is built up efficiently ($p_z \gg p_\perp$) the dark gauge boson production in the unbroken phase is given by Eq.~\eqref{eq:EBm0}. All charged SM leptons contribute to the induced current. For simplicity we will take $B \sim E$ in the estimates below. This enters in the estimate of the typical fermion energy at production, $\bar \omega$, and moreover constrains $\beta \lesssim 1$ in Eq.~\eqref{eq:EBm0} with $\beta < 1$ indicating that the upper bound on $EB$ is not saturated. With this, the (would-be) fermion temperature is obtained as
\begin{align}
	T_\psi =  6 \cdot 10^3 \, H \,  \beta^{5/16} \left( \frac{\xi}{10} \right)^{5/8} \left( \frac{0.01}{\theta_D} \right)^{5/4} \,,
\end{align}
indicating that thermalisation in the fermion sector is avoided if
\begin{align}
	\Gamma_\text{th} \simeq \alpha^2 \, T_\psi < H \quad \Leftrightarrow \quad \theta_D \gtrsim 4 \cdot 10^{-3} \,,
	\label{eq:thermal_regime_m0}
\end{align}
for the reference values $\xi = 10$ and $\beta = 1$. 
From Eq.~\eqref{eq:Gamma_sc} we obtain the scattering rate for a non-thermal distribution as
\begin{align}
	\Gamma_\text{sc} \simeq  0.4 \,  H \, \beta^{1/2} \left( \frac{\xi}{10} \right) \left( \frac{0.01}{\theta_D} \right)^2 \,,
\end{align}
which yields a similar conclusion regarding the value $\theta_D$ required for thermalisation in the fermion sector.

\paragraph{Induced current.} For non-thermal fermions, i.e.\ $\tau = \Gamma_{sc}^{-1} \gtrsim H^{-1}$, the ratio of momenta parallel and perpendicular to the $z$-axis is
\begin{align}
	\frac{p_z}{p_\perp} \simeq \frac{e \theta_D E \Gamma_\text{sc}^{-1}}{\bar \omega} \simeq 10^4 \, \beta^{-1/4} \, \left( \frac{10}{\xi} \right)^{1/2} \left( \frac{\theta_D}{0.01} \right)\,.
\end{align}
Inserting Eq.~\eqref{eq:thermal_regime_m0} we see that $p_z/p_\perp \gg 1$ for 
\begin{align}
 \theta_D \gtrsim 10^{-6} \beta^{1/4} \left(\frac{\xi}{10}\right)^{1/2}\,.
\end{align}
Hence the expression~\eqref{eq:Jind} for the induced current, and consequently the expression~\eqref{eq:EBm0} for the dark photon production is valid in this regime.
For thermalised fermions, $\tau = \Gamma_\text{th}^{-1}$, we obtain
\begin{align}
	\frac{p_z}{p_\perp} \simeq \frac{e \theta_D E \Gamma_\text{th}^{-1}}{T_\psi} \simeq 3 \cdot 10^4 \, \beta^{-1/8} \, \left( \frac{10}{\xi} \right)^{1/4} \left( \frac{\theta_D}{0.01} \right)^{1/2}\,.
\end{align}
In summary,  the induced current builds up efficiently as long as the Higgs vev is zero and consequently Eqs.~\eqref{eq:Jind} and \eqref{eq:EBm0} are valid as long as

\begin{align}
	\theta_D \gtrsim  10^{-11} \beta^{1/4} \left( \frac{\xi}{10} \right)^{1/2} \,.
	\label{eq:th-1}
\end{align}

\paragraph{Thermalisation within the dark gauge boson sector.} Turning to the thermalisation among the dark gauge bosons, Eq.~\eqref{eq:TX} yields
\begin{align}
	\frac{\Gamma_X}{H} \simeq  8 \cdot 10^{-4} \, \beta^{1/4}  \left( \frac{\xi}{10} \right)^{1/2} \left( \frac{\theta_D}{0.01} \right)^{1/2} \ll 1 \,,
	\label{eq:Xthermalisation}
\end{align}
indicating that the dark gauge bosons remain non-thermal in the entire parameter space of interest covered by Eq.~\eqref{eq:th-1}.

\paragraph{Thermalisation within the SM photon sector.} The induced current provides a source term for a non-thermal photon field configuration. As long as these photons do not thermalise, we expect Schwinger production of leptons to restrict the amplitude of this gauge field configuration according to Eq.~\eqref{eq:EBm0} with $\theta_D = 1$. Taking the limit $\theta_D = 1$ in Eq.~\eqref{eq:Xthermalisation}, we obtain
\begin{equation}
  8 \cdot 10^{-3} \, \beta^{1/4}  \left( \frac{\xi}{10} \right)^{1/2}  \ll 1 
\end{equation}
as a condition to avoid efficient scattering for the non-thermal SM photon population.

\subsection{Broken phase}

As long as the induced current is built up efficiently, the dark photon gauge boson production in the broken phase is given by Eq.~\eqref{eq:EB}. We will assume here that only the electrons are efficiently produced,  though for sufficiently large values of the electric field the heavier fermions species will gradually become important, too. As above, we will assume $E \sim B$ for simplicity. 

In the regime where the exponential suppression of the fermion production due to the electron mass is mild, $z \ll 1$, we recover the expressions derived for the unbroken phase with the replacements
\begin{align}
	c \mapsto \tilde c \,, \quad \beta \mapsto \tilde \beta \,.
\end{align}
With this, we find that the induced current builds up efficiently for
\begin{align}
	\theta_D \gtrsim 10^{-6} \tilde{\beta}^{1/4} \left( \frac{\xi}{10} \right)^{1/2} \,,
\end{align}
and dark gauge boson thermalisation in negligible in the entire parameter space of interest.

Next let us consider the regime $z > 1$. In this case, $W_0(z) \simeq \ln(z)$, and we thus parametrize $EB/H^4 = \tilde b d^2$ with $\tilde b = \tilde \beta \ln(z) \lesssim 10$ for $z < 10^4$. With this, we obtain
\begin{align}
	T_\psi \simeq  10^4 \; \tilde b^{5/16} \, e^{-1/(4 \sqrt{\tilde b})} H z^{5/8} \left( \frac{\xi}{10} \right)^{5/8} \left( \frac{0.01}{\theta_D} \right)^{5/4} \,,
	\label{eq:Tpsi_broken}
\end{align}
and
\begin{align}
	\frac{\Gamma_\text{th}}{H}  \simeq 0.6  \; \tilde b^{5/16} \, e^{-1/(4 \sqrt{\tilde b})} z^{5/8} \left( \frac{\xi}{10} \right)^{5/8} \left( \frac{0.01}{\theta_D} \right)^{5/4} \,,
	\label{eq:thermalisation_broken}
\end{align}
which indicates that fermion thermalisation is efficient in most of the parameter space.
Turning to the induced current, this thermal fermion population yields
\begin{align}
	\frac{p_z}{p_\perp} \simeq \frac{p_z}{T_\psi} \simeq 2 \cdot 10^4 \; \tilde b^{-1/8}  \, e^{-1/(4 \sqrt{\tilde b})} z^{-1/4} \left( \frac{10}{\xi} \right)^{1/4} \left( \frac{\theta_D}{0.01} \right)^{1/2} \,,
\end{align}
implying that the induced current builds up efficiently as long as
\begin{align}
	\theta_D \gtrsim 2 \cdot 10^{-11} \; \tilde b^{1/4} \; e^{-1/ \sqrt{\tilde b}} \, z^{1/2} \left( \frac{\xi}{10} \right)^{1/2}  \,.
\end{align}
Hence, the induced current builds up efficiently also once the Higgs acquires its vev.

Finally, considering the thermalisation within the dark gauge boson sector for $z > 1$, we find
\begin{align}
\frac{\Gamma_X}{H} =  10^{-7} \, \tilde b^{1/4} z^{1/2} \theta_D^{1/2} \xi^{1/2} \ll 1 \,,
\end{align}
indicating that dark gauge bosons do not scatter efficiently. Taking the limit $\theta_D = 1$, we note that this conclusion also holds for the SM photons.

\section{Thermal dark relics}
\label{app:thermaldarkrelics}

Once the SM sector reheats, dark sector quanta (relaxion particles and dark photons) are thermally produced through their (small) interactions with the SM plasma. Note that the fully diagonalised basis~\eqref{eq:basis1} has no interaction term between the relaxion and the SM, whereas in the interaction basis~\eqref{eq:interactionbasis} in the limit $m_X \rightarrow 0$, the only coupling of the dark photon to the SM involves the relaxion. Consequently, omitting unphysical effects which can be rotated away in a suitable basis, the leading-order interaction process between the SM and the dark sector involves both the dark photon and the relaxion, $e^+ e^- \rightarrow A' \rightarrow X \phi$. 


In the limit $m_X \ll m_\phi <  E$ with $E$ denoting the center of mass energy of this $2 \rightarrow 2$ process, the relevant thermally averaged cross section is given by~\cite{Hook:2021ous}
\begin{align}
 \sigma_{e^+ e^- \rightarrow \phi X} = \frac{\alpha_D^2 \theta_D^2 \alpha_\text{em}}{24 \pi^2 f_D^2}  
 \left[ \frac{5}{2} - \frac{3 m_\phi^2}{8 T^2}  + {\cal O}\left(\frac{m_\phi^4}{T^4}\right) \right]\,,
 \label{eq:B1}
\end{align}
and the number density of electrons is given by the thermal equilibrium value, $n_e = 2 /\pi^2 \, T^3$ for $T \gg m_e$. 
The dark sector decouples when the rate of this process becomes small compared to the Hubble expansion rate, with
\begin{align}
\frac{\Gamma_{e^+ e^- \rightarrow \phi X}}{H}&  = \frac{n_e M_*  \sigma_{e^+ e^- \rightarrow \phi X}}{T^2} \nonumber \\
& \simeq 10^{-3} \; g_{*}^{-1/2} \left( \frac{\theta_D}{10^{-3}}\right)^2  \left( \frac{10^4~\text{GeV}}{f_D/\alpha_D} \right)^2  \left(\frac{T}{\text{MeV}}\right)  \left( 1 - \frac{3 m_\phi^2}{20 T^2} \right) \,,
 \end{align}
which defines the decoupling temperature $T_\text{dec,d}$ as the temperature when $\Gamma/H = 1$. 
If this interaction stays efficient until below the QCD phase transition,  the resulting energy density in the dark sector is comparable to that of the visible sector, in contradiction with the constraints from $N_\text{eff}$. For $m_\phi > m_e$, the relaxion production ceases at $T_\text{dec,d} \gtrsim m_\phi > m_e$. Assuming  $T_\text{dec,d} >$~MeV, which is the decoupling temperature of the neutrinos, the radiation stored in the decoupled dark sector can be parametrised in terms of additional relativistic neutrino species,
\begin{align}
 \rho_\text{rad}(T) = \rho_\text{rad}^\text{SM}(T) + \rho_\text{rad}^d(T)  = \frac{\pi^2}{30} \left[ 2 + \frac{7}{4} \left( \frac{4}{11} \right)^{4/3} (N_\text{eff}^\text{SM} + \Delta N_\text{eff}) \right] T^4 \,,
\end{align}
with $N_\text{eff}^\text{SM} = 3.044$ and 
\begin{equation}
 \Delta N_\text{eff} = \frac{16}{7} \left( \frac{43}{4 g_{*,s}(T)} \right)^{4/3} \,,
\end{equation}
where 43/4 denotes the relativistic SM degrees of freedom  just before neutrino decoupling, i.e.\ photons, neutrinos and electrons.
Requiring $\Delta N_\text{eff} \lesssim 0.5$~\cite{Pisanti:2020efz,Yeh:2020mgl} requires $g_{*,s}(T) \gtrsim 30$ and hence a decoupling temperature above the QCD phase transition, $T_\text{dec,d} \gtrsim 100$~MeV~\cite{Kolb:1990vq}.

\section{Plasma mass}
\label{app:inmedium}

In the SM thermal plasma, the SM photon obtains a plasma mass $m_\gamma \simeq 0.1~T$~\cite{Braaten:1993jw} in the interaction basis, Eq.~\eqref{eq:interactionbasis}. In our scenario, this becomes relevant for the thermal dark relic production in the SM plasma after reheating and in the astrophysical environments relevant for the stellar cooling and supernova contraints. These computations are performed directly in the interaction basis and the plasmon mass is directly included. For completeness, we give here the basis transformation to recover the diagonal basis~\eqref{eq:basis1} in which the kinetic terms are canonical and the mass matrix is diagonal.
Following the same procedure as in the previous subsection, we obtain the rotation matrix
\begin{align} R = \begin{pmatrix} 1  & \frac{\epsilon \theta_D}{1 - \epsilon} \\
\frac{\epsilon \theta_D}{1 - \epsilon} & -1  \end{pmatrix}  + {\cal O}(\theta_D^2) \,,
\end{align}  
where $\epsilon \equiv (m_X/m_\gamma)^2 \ll 1$. In this basis we then obtain
\begin{align}
 & m_A^2 = m_\gamma^2 \,, \quad m_X^2 = m_{X'}^2 \,, \quad \frac{\alpha_D}{f} X^{\mu \nu} \tilde X_{\mu \nu} \mapsto  \frac{\alpha_D}{f} ( X^{\mu \nu} \tilde X_{\mu \nu} -  \frac{2 \theta_D}{1 - \epsilon} X^{\mu \nu} \tilde A_{\mu \nu} ) \,, \nonumber \\
 & e Q_i (A_\mu - \theta_D  X_\mu) \mapsto e Q_i (A_\mu - \theta_D \frac{\epsilon}{-1 +\epsilon} X_\mu)\,. 
\end{align}
From the last line we see that the introduction of a mass hierarchy in the interaction basis suppresses the effective fermion millicharge by a factor $\epsilon$. On the other hand, from the last term in the first line we note an additional coupling between the dark and the visible photon which vanishes in the limit $m_\gamma \rightarrow 0$.

\addcontentsline{toc}{section}{References}
\bibliographystyle{JHEP}
\bibliography{refs_final}

\end{document}